\newcommand{\changev}{}

\documentclass[traditabstract]{aa} 
\usepackage{amsmath}
\usepackage{multirow}
\usepackage{graphicx}
\usepackage{txfonts}
\usepackage{amsfonts}
\usepackage{subfigure}
\usepackage{amssymb,amsfonts}
\usepackage{natbib}
\bibpunct{(}{)}{;}{a}{}{,}

\begin{document}

\title{Inner disk clearing around the Herbig Ae star HD\,139614: Evidence for a planet-induced gap ?\thanks{Based on observations collected at the European Southern Observatory, Chile (ESO IDs : 385.C-0886, 087.C-0811, 089.C-0456, and 190.C-0963).}}

\author{A. Matter\inst{1,2}\fnmsep\thanks{Corresponding author : Alexis.Matter@oca.eu}\thanks{Present address: Observatoire de la C\^ote d'Azur, boulevard de l'Observatoire, CS 34229, 06304 Nice, France.} \and L. Labadie\inst{3} \and J. C. Augereau\inst{1} \and J. Kluska\inst{4} \and A. Crida\inst{3,5} \and A. Carmona\inst{6,7} \and {J.F. Gonzalez}\inst{8}  \and W. F. Thi\inst{9} \and J.-B. Le Bouquin\inst{1} \and J. Olofsson\inst{10} \and B. Lopez\inst{2}}

\institute{
Univ. Grenoble Alpes, IPAG, F-38000 Grenoble, France; CNRS, IPAG, F-38000 Grenoble, France. \and Laboratoire Lagrange, Universit\'e C\^ote d'Azur, Observatoire de la C\^ote d'Azur, CNRS, Boulevard de l'Observatoire, CS 34229, 06304 Nice Cedex 4, France.
 \and I. Physikalisches Institut, Universit\"at zu K\"oln, Z\"ulpicher Str. 77, 50937 K\"oln, Germany. \and University of Exeter, School of Physics, Stocker Road, Exeter, EX4 4QL, UK. \and Institut Universitaire de France, 103 boulevard Saint Michel, F-75005 Paris, France. \and Universidad Aut\'onoma de Madrid, Campus Cantoblanco, 28049 Madrid, Spain. \and Konkoly Observatory, Hungarian Academy of Sciences, H-1121 Budapest, Konkoly Thege Mikl\'os \'ut 15-17, Hungary.  \and Univ. Lyon 1, Observatoire de Lyon, 9 avenue Charles Andr\'e, F-69230 Saint-Genis Laval, France; CNRS, CRAL, F-69230 Saint-Genis Laval, France; Ecole Normale Sup\'erieure de Lyon, F-69007 Lyon, France. \and Max Planck Institut f\"ur extraterrestrische Physik, Giessenbachstrasse 1, 85748 Garching, Germany.  \and Max Planck Institut f\"ur Astronomie, K\"onigstuhl 17, D-69117 Heidelberg, Germany. 
 }

\abstract  
{Spatially resolving the inner dust cavity (or gap) of the so-called (pre-)transitional disks is a key to understanding the connection between the processes of planetary formation and disk dispersal. The disk around the Herbig star HD\,139614 is of particular interest since it presents a pretransitional nature with an au-sized gap structure that is spatially resolved by mid-infrared interferometry in the dust distribution.
 With the aid of new near-infrared interferometric observations, we aim to characterize the 0.1--10~au region of the HD~139614 disk further and then identify viable mechanisms for the inner disk clearing.\\
We report the first multiwavelength modeling of the interferometric data acquired on HD~139614 with the VLTI instruments PIONIER, AMBER, and MIDI, complemented by Herschel/PACS photometric measurements. We first performed a geometrical modeling of the new near-infrared interferometric data, followed by radiative transfer modeling of the complete dataset using the code RADMC3D.\\
We confirm the presence of a gap structure in the warm $\mu$m-sized dust distribution, extending from about 2.5~au to 6~au, and constrained the properties of the inner dust component: e.g., a radially increasing dust surface density profile, and a depletion in dust of $\sim 10^3$ relative to the outer disk. 
 Since self-shadowing and photoevaporation appears unlikely to be responsible for the au-sized gap of HD~139614, we thus tested if dynamical clearing could be a viable mechanism using hydrodynamical simulations to predict the structure of the gaseous disk. Indeed, a narrow au-sized gap is consistent with the expected effect of the interaction between a single giant planet and the disk. 
 Assuming that small dust grains are well coupled to the gas, we found that an approximately 3~$M_{\rm jup}$ planet located at $\sim 4.5$~au from the star could, in less than 1~Myr, reproduce most of the aspects of the dust surface density profile, while no significant depletion (in gas) occurred in the inner disk, in contrast to the dust. However, this "dust-depleted" inner disk could be explained by the expected dust filtration by the gap and the efficient dust growth/fragmentation occurring in the inner disk regions.\\
Our results support the hypothesis of a giant planet opening a gap and shaping the inner region of the HD~139614 disk. This makes HD~139614 an exciting candidate specifically for witnessing planet-disk interaction.}


\authorrunning{A. Matter et al.}
\titlerunning{Interferometric view on HD\,139614}

\maketitle
%

\section{Introduction}
Viscous accretion, photoevaporation, and dynamical clearing are processes through which disks are thought to shape and dissipate most of their mass, which lies in the gas content \citep[see][ for a review]{2011ARA&A..49...67W,2014prpl.conf..667B,2014prpl.conf..475A}. Dust also
plays an important role since it dominates the disk opacity and provides the raw material for building the rocky planets and the giant-planet cores. Dust is affected by processes stemming from its coupling to the gas such as dust trapping, radial migration, and grain growth \citep[e.g., ][]{2012A&A...539A.148B,2014MNRAS.437.3037L}, or from the stellar irradiation \citep[e.g., ][]{2011A&A...531A.101D}. Such evolution processes can produce specific signatures such as gaps, inner holes, and asymmetries \citep[see, e.g.,][]{2007MNRAS.377.1324C,2011MNRAS.412...13O,2012A&A...545A.134M}. 
Identifying them in the inner disk regions ($\sim 0.1-10$~au) is essential since these regions are the expected cradle of telluric planets \citep{Righter27062011} and the location of the photoevaporation onset \citep{2014prpl.conf..475A}. The emission deficit in the infrared (IR) SED of pre-transitional and transitional disks \citep[e.g., ][]{1989AJ.....97.1451S,2010ApJ...712..925C,2010ApJ...718.1200M} has been commonly interpreted as a clearing of their inner regions, which make these objects relevant laboratories for observing the signatures of disk evolution and dissipation processes.
Infrared interferometry can specifically probe these inner disk regions \citep{2010ARA&A..48..205D,Carmona2014} and detect dust radial evolution \citep[][]{2004Natur.432..479V}, brigthness asymmetries \citep[e.g., ][]{2013ApJ...768...80K}, and dust clearing \citep[e.g.,][]{2010A&A...511A..75B,2014A&A...564A..93M}. \\
\noindent The Herbig star HD\,139614 (see Table~\ref{tab:star}) is of particular interest here.
It is associated with the Upper Centaurus Lupus (UCL) region of the Sco OB2 association \changev{located at $140\pm27$~pc \citep[][ and references therein]{2008hsf2.book..235P}}. 
Its group-I SED \citep{2001A&A...365..476M} presents pretransitional features (near-infrared (NIR) excess with mid-infrared (MIR) emission deficit at 6--7~$\mu$m) and weak MIR amorphous silicate features \citep{2010ApJ...721..431J}, which suggests significant dust evolution in the inner regions. 
High-resolution imaging with T-RECS/Gemini only resolved the MIR continuum emission at 18~$\mu$m \citep[17$\pm$4~au, ][]{2011ApJ...737...57M}, while MIR interferometry probed the inner N-band emitting region and revealed a narrow gap-like structure (located from $\sim$~2.5 to 6~au) depleted in warm $\mu$m-sized dust \citep{2014A&A...561A..26M}.
Recent works showed or found hints of the presence of gaps in Group I disks \citep[e.g., ][]{2013A&A...555A..64M,2015A&A...581A.107M}, and HD\,139614 is a rare, if not unique, case of object for which a small au-sized gap ($\sim 3$~au) has been spatially resolved.
Previous studies, mainly based on sub-mm interferometry \citep[e.g., ][]{2011ApJ...732...42A,2014ApJ...783L..13P} and NIR imaging \citep{2013ApJ...766L...2Q,2014ApJ...790...56A}, have focused on objects with large cavities ($\sim 10$--100~au). Their origin is ambiguous and possibly combines, for example, photoevaporation \citep{2013MNRAS.430.1392R}, magnetorotational instability \citep{2007NatPh...3..604C}, and/or multiple unseen planets \citep{2011ApJ...729...47Z}. Single Jovian planets are expected to open au-sized gaps while inducing a gas and dust surface density decrease in the inner disk regions \citep[e.g., ][]{2007MNRAS.377.1324C}. Knowing that spectroscopic observations have proven the presence of gas \citep{2009A&A...508..707P,2012A&A...544A..78M} and gas tracers like PAHs \citep{2010ApJ...718..558A} in the HD~139614 disk, this object constitutes a unique opportunity to characterize the early stages of inner disk dispersal and \changev{potentially} witness planet-disk interaction. 
 Spatially resolved IR observations are thus required.\\ 
\noindent We report the first multiwavelength analysis of HD\,139614, combining MIR VLTI/MIDI data with  
new NIR interferometric data obtained with VLTI/PIONIER \citep{2011A&A...535A..67L} and VLTI/AMBER \citep{2007A&A...464....1P}, and far-IR photometry with Herschel\footnote{Herschel is an ESA space observatory with science instruments provided by
European-led Principal Investigator consortia and with important
participation from NASA.}/PACS \citep{2010A&A...518L...1P,2010A&A...518L...2P}.  We aim to obtain new and robust constraints on the innermost $\sim 0.1$--1~au dust structure. We also aim to use radiative transfer to refine the constraints of the previous analytical modeling of the MIDI data (e.g., gap characteristics, outer disk properties). This
is essential to determining the degree of differentiation of the inner and outer disk regions, on either side of the gap, and identify viable mechanisms for the inner disk clearing around HD~139614.\\ 
Section 2 summarizes the new observations that complement the previous VLTI/MIDI and optical/IR SED data used in \citet{2014A&A...561A..26M}. Section 3 presents the analysis and geometrical modeling of the PIONIER and AMBER data. Section 4 describes the radiative transfer modeling of the broadband SED and the complete set of interferometric data. Section 5 discusses the modeling results against the mechanisms possibly responsible for the gap structure. This includes a comparative study with hydrodynamical simulations of gap opening by a planet. Finally, Sect. 6 summarizes
our work and outlines the perspectives.
\begin{table*}
\caption{Stellar parameters used for HD\,139614}
 \centering
 \begin{tabular}{cccccccc}
 \hline
 \hline
{\footnotesize $d^{(1,2)}$ [pc]}& {\footnotesize $A_{\rm V}^{(3)}$ [mag]}&{\footnotesize SpTyp$^{(4)}$}&{\footnotesize $M_*^{(4)}$ [M$_{\sun}$]}&{\footnotesize $R_*^{(4)}$ [R$_{\sun}$]}&{\footnotesize $\log T_*^{(4)}$ [K]}&{\footnotesize $\log g^{(5)}$}&{\footnotesize Age$^{(6)}$ [Myr]}\\
 \hline
{\footnotesize 140$\pm 27$}&{\footnotesize 0.09 }&{\footnotesize A7V}&{\footnotesize 1.7$\pm 0.3$}&{\footnotesize 1.6}&{\footnotesize 3.895}&{\footnotesize 4.0}&{\footnotesize $8.8_{-1.9}^{+4.5}$ }\\      
 \hline
 \end{tabular}
 \tablefoot{The uncertainties are taken from the references given below. The stellar radius $R_*$ was derived from the stellar luminosity and $T_*$ taken from \citet{2005A&A...437..189V}; no uncertainty is available, but it is consistent with other radius estimates \citep[e.g.,][]{2013MNRAS.429.1001A}. The $\log g$ and $M_*$ values are consistent with our derived $R_*$ value \citep[see, e.g.,][]{2013MNRAS.429.1001A}.}
 \tablebib{(1): \citet{1999AJ....117..354D}; (2): \citet{2008hsf2.book..235P}; (3): \citet{1999A&A...345..547Y}; (4): \citet{2005A&A...437..189V};  (5): \citet{1997MNRAS.289..831S}; (6): \citet{2013MNRAS.429.1001A}. }
 \label{tab:star}
 \end{table*}
 
\section{Observations and data processing}
\begin{table*}[t]
\caption{Observing log of 
HD\,139614.}
\centering
 \begin{tabular}{ccccccc}
 \hline
 \hline
{\footnotesize Date}&{\footnotesize UT}&{\footnotesize Baseline}&{\footnotesize Calibrator}&{\footnotesize Seeing (\arcsec)}&{\footnotesize Airmass}&{\footnotesize Label}\\
 \hline
 \multicolumn{7}{c}{PIONIER}\\
 \hline
 {\footnotesize 25/03/2012}& {\footnotesize 08:55:41}&{\footnotesize A1-G1-K0-I1}&{\footnotesize HD\,145191} & {\footnotesize 1.3}&{\footnotesize 1.0}& {\footnotesize N/A}  \\ 
 {\footnotesize 06/06/2013}& {\footnotesize 01:27:00}&{\footnotesize A1-G1-J3-K0} &{\footnotesize HD\,137598} & {\footnotesize 1.1}&{\footnotesize 1.1}& {\footnotesize N/A} \\ 
 &&&{\footnotesize HD\,141702} &&& \\
 {\footnotesize 16/06/2013}& {\footnotesize 23:31:00}&{\footnotesize D0-G1-H0-I1} &{\footnotesize HD\,137598} & {\footnotesize 1.0}&{\footnotesize 1.3}&{\footnotesize N/A} \\
 {\footnotesize 03/07/2013}& {\footnotesize 23:58:00}&{\footnotesize A1-B2-C1-D0} &{\footnotesize HD\,137598} & {\footnotesize 1.1}&{\footnotesize 1.1}&{\footnotesize N/A} \\
 &&&{\footnotesize HD\,141702} &&& \\ 
 \hline
 \multicolumn{7}{c}{AMBER}\\
 \hline 
 {\footnotesize 09/05/2012}& {\footnotesize 05:27:00}&{\footnotesize UT1-UT2-UT4} &{\footnotesize HD\,141702}&{\footnotesize 1.0}&{\footnotesize 1.0}&{\footnotesize 1,2,3}       \\ 
 {\footnotesize 09/05/2012}& {\footnotesize 06:27:00}&{\footnotesize UT1-UT2-UT4} &{\footnotesize HD\,140785}&{\footnotesize 0.6}&{\footnotesize 1.1}&{\footnotesize 4,5,6}     \\ 
 \hline
 \end{tabular}
 \tablefoot{ Each calibrated observation lasted 45~min to 1~h. The airmass and seeing at $\lambda=0.5$~$\mu$m are the mean values of each observation. The last column gives a label for the three UV points of each AMBER baseline triplet.} 
 \label{tab:log}
 \end{table*}

\subsection{VLTI/PIONIER observations}  
We observed HD\,139614 with the 1.8~m Auxiliary Telescopes (AT) in the frame of a large program on Herbig stars (ID: 190.C-0963) conducted with VLTI/PIONIER (see Table~\ref{tab:log}). It combines the light from four telescopes in {\itshape H} band, and provides six squared visibilities, noted as $V^2$, and four closure phases per ATs configuration. Our observations were obtained at low spectral resolution ($R\sim 40$) providing three spectral channels centered at 1.58~$\mu$m, 1.67~$\mu$m, and 1.76~$\mu$m.
\changev{The calibration stars were selected with the SearchCal tool from the Jean-Marie Mariotti Center (JMMC). Each HD~139614 observation was bracketed by two calibrator observations, and the estimated transfer function was interpolated at the time of observation.}
The projected baseline lengths range from 10.3~m to 139.8~m ($\lambda/2B \simeq 17$~mas to 1~mas). The PIONIER field of view (FOV) is $\sim 200$~mas ($\sim 25$~au at 140~pc).  
We performed a standard data reduction using the pipeline
"pndrs", described in \citet{2011A&A...535A..67L}. Figure~\ref{fig:uv} shows
the UV coverage and calibrated data. \changev{The final uncertainties include the statistical errors and the uncertainty on the transfer function.}


 
 \subsection{VLTI/AMBER observations}
\begin{figure*}[t]
 \centering
 \resizebox{\hsize}{!}{\includegraphics[scale=0.1]{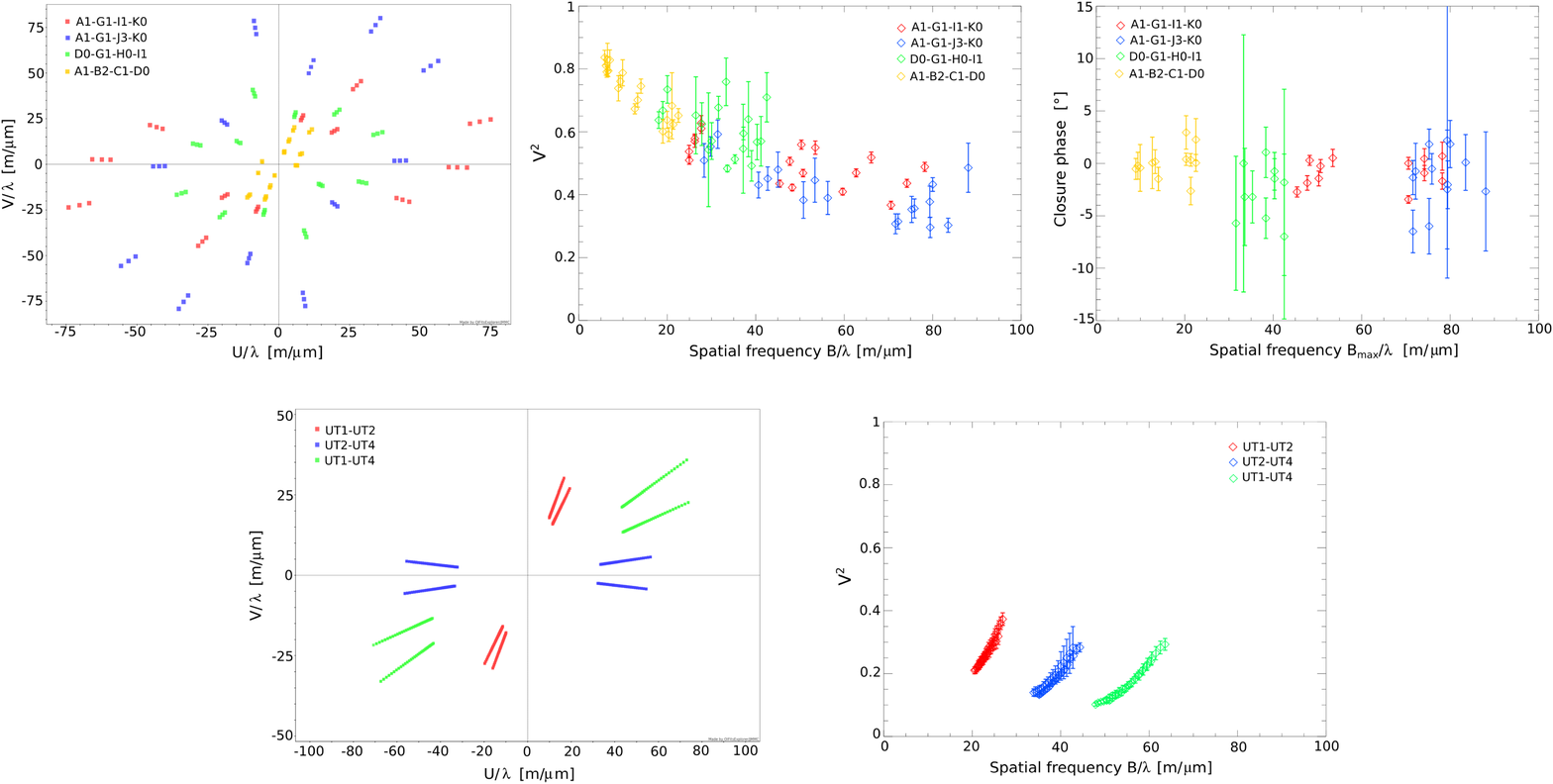}}
 \caption{{\footnotesize Top (from left to right): (u, v) coverage, PIONIER $V^{2}$ and closure phases; for a given quadruplet, every baseline "observation" consists of 3 measurements at 1.58~$\mu$m, 1.67~$\mu$m, and 1.76~$\mu$m; For each closure phase, $B_{\rm max}$ is the projected length of the longest baseline of the triplet. Bottom (from left to right): (u, v) coverage, and AMBER {\itshape K}-band $V^{2}$; every baseline "observation" consists of $V^2$ measurements between 2~$\mu$m and 2.5~$\mu$m.}}
 \label{fig:uv}
\end{figure*}

HD\,139614 was observed with AMBER (program 0.89.C-0456(A)) using the 8-m telescope triplet UT1-UT2-UT4. AMBER can combine three beams and provides spectrally dispersed $V^2$ and closure phases. Our observations were obtained in low spectral resolution ($R\sim 35$) in the {\itshape H} and {\itshape K} bands and included two interferometric calibrators (see Table~\ref{tab:log}) chosen using the SearchCal tool. 
The data consisted of two sets of eight and five exposures of 1000 frames.
The data were reduced with the JMMC "amdlib" package (release 3.0.5). 
For each exposure, a frame selection was made to minimize the impact of the instrumental jitter and the non-optimal light injection into the optical fibers. Twenty percent of the frames with the highest fringe S/N provided the smallest errors on the resulting $V^2$. Despite this, the {\itshape H}-band data showed significant variability and low S/N ($\sim 1.5$) probably because of the {\itshape H}$=7.3$ magnitude of HD\,139614 that was close to the AMBER limiting magnitude ({\itshape H}$_{\rm corr}=7.5$). Moreover, the low-resolution closure phases are affected by a strong dependency on the piston, which is not reduced by stacking frames (see AMBER manual, ESO doc. {\small VLT-MAN-ESO-15830-3522}). Therefore, from the the AMBER observations, we only kept the {\itshape K}-band $V^2$ measurements.
The instrumental transfer function was calibrated using the closest calibrator in time. 
The calibrated $V^2$ errors include the statistical error obtained when averaging the individual frames and the standard deviation of the transfer function over the calibrator observations. Our final dataset consists of six dispersed $V^2$ in the [2.0--2.5]~$\mu$m range. The projected baseline lengths range from 52.0~m to 127.7~m (5~mas to 1.8~mas at $\lambda=2.2$~$\mu$m). The AMBER FOV is $\sim 60$~mas ($\sim 9$~au at 140~pc). Figure~\ref{fig:uv} shows the UV coverage and calibrated data. 

\subsection{Photometric data}
 
We complemented the SED used in \citet{2014A&A...561A..26M} with {\itshape Herschel}/PACS observations acquired on 7 March 2011 in the frame of the Key Program GASPS \citep[][]{2013PASP..125..477D}. Data reduction, flux extraction, and error estimation were performed as in \citet{2013A&A...551A.134O}, and lead to $(9.66\pm0.96)\times10^{-13}$~W.m$^{-2}$ at 70~$\mu$m, $(4.93\pm0.01)\times10^{-13}$~W.m$^{-2}$ at 100~$\mu$m, and $(2.43\pm0.01)\times10^{-13}$~W.m$^{-2}$ at 160~$\mu$m. We also included measurements at 800~$\mu$m and 1.1~mm taken from \citet{1996MNRAS.279..915S}. 
Except for the recent Herschel data, we assumed a 10\% relative uncertainty on the SED points, which is conservative, especially for the 2MASS data with formal uncertainties $\sim$ 5\% \citep[][]{2006AJ....131.1163S}. The broadband SED is shown in Fig.\ref{fig:radmcsed}.

\subsection{Stellar parameters and spectrum}
Table~\ref{tab:star} shows the stellar parameters used for HD~139614. Following \citet{2008hsf2.book..235P}, we adopt a distance of $140\pm 27$~pc hereafter. For the stellar flux, we use the same Kurucz spectrum ($T_{\rm eff}=7750$~K, $\log g=4.0$, Fe/H=-0.5) as in \citet{2014A&A...561A..26M}. Given the temperature quoted in Table~1 ($T_{\rm eff}\sim7850$~K), this may be slightly too cold but remains consistent with other temperature estimates \citep[e.g., ][]{2013MNRAS.429.1001A}. 

\section{Observational analysis and geometrical modeling}

\subsection{Observational results}
As shown in Fig.~\ref{fig:uv}, the PIONIER (top) and AMBER (bottom) $V^2$ shows circumstellar emission that is well resolved at a level of a few mas.
The PIONIER $V^2$ data present an exponential-like profile with a steep decrease at low frequencies ($\leq 10$~m/$\mu$m) that may suggest a fully resolved emission. The PIONIER FOV ($\sim 25$~au) partly encompasses the outer disk, which starts at $\sim 6$~au \citep{2014A&A...561A..26M}. NIR scattering by sub- and micron-sized grains at the outer disk's surface may thus contribute to this steep decrease. At high frequencies ($\geq 40$~m/$\mu$m), the $V^2$ reach an asymptotic level between 0.5 at 1.58~$\mu$m and 0.35 at 1.76~$\mu$m, which translates to a visibility level between 0.7 and 0.6, respectively. This is close to the stellar-to-total flux ratio (STFR) evaluated to 0.7 at 1.65~$\mu$m using the stellar Kurucz spectrum of Sect.~2.5 and the NIR SED (2MASS measurements). This indicates that the circumstellar emission is fully resolved by PIONIER at high spatial frequencies. In the {\itshape K} band, the AMBER $V^2$ measurements range from 0.1 at 2~$\mu$m to 0.3 at 2.5~$\mu$m for the longest baselines (UT2-UT4 and UT1-UT4) and from 0.15 to 0.35 for the shortest one (UT1-UT2). The corresponding average $V$ level is 0.4 (UT2-UT4 and UT1-UT4) and 0.5 (UT1-UT2) across the {\itshape K} band, which is close to the STFR evaluated to 0.4 at 2.2~$\mu$m. The inner disk is thus fully resolved in the {\itshape K} band, at least for the longest baselines. This suggests a NIR-emitting region that probably spreads over at least one au ($\sim 7$~mas at 140~pc), as already suggested by \citet{2014A&A...561A..26M}.\\
Our observations also show that $V^2$ depends on wavelength across both spectral bands. Such a chromatic effect has been studied, for instance, in the frame of IR image reconstruction \citep{2014A&A...564A..80K}. At high spatial frequencies, where the disk is fully resolved, the chromatic variation in the asymptotic level of visibility is $\sim 0.1$ (expressed in $V$) across the {\itshape H} band and $\sim 0.2$ across the {\itshape K} band. This reveals the decrease in the unresolved stellar contribution combined with a positive chromatic slope due to the emission from the hottest dust grains. Assuming a blackbody law for this hottest component, we could reproduce the NIR SED (from $\lambda=1.25$~$\mu$m to $\lambda=3.4$~$\mu$m) and the STFR chromaticity across the {\itshape H} band ($\sim 0.1$) and the {\itshape K} band ($\sim 0.2$), with dust grain temperatures from 1300 to 1500~K, which is close to the sublimation temperature for silicates. We note that the hottest emission at 1500~K induces a slightly lower STFR ($\sim 0.65$) at 1.65~$\mu$m than the ratio induced by the blackbody emission at 1300~K ($\sim 0.7$). The latter is closer to the STFR derived from the 2MASS measurement at 1.65~$\mu$m, namely $0.7$.\\ 
The PIONIER closure phases do not show a clear departure from zero, hence noticeable signatures of brightness asymmetries. We do note a dispersion of $\lesssim 2^{\circ}$ for the yellow and red measurements and $\lesssim 5^{\circ}$ for the noisier blue and green ones (see Fig.\ref{fig:uv}). Since the closure phase produced by a binary system is, at the first order, proportional to the flux ratio between the components \citep{2006MNRAS.367..825V}, a closure phase dispersion of $\pm 2-5^{\circ}$ would translate to an upper limit of $\sim 3-8\times10^{-2}$ on the flux ratio. A flux ratio of $\sim 10^{-2}$ would imply an upper mass limit for the hypothetical companion of about 0.11~M$_{\odot}$, using the recent BT-SETTL atmospheric models for very low-mass stars, brown dwarfs, and exoplanets of \citet{2012RSPTA.370.2765A}. Based on the available PIONIER data, it is thus unlikely that HD\,139614 is actually a tight binary system hosting a companion that is more massive than 0.11~M$_{\odot}$ in its nearby environment ($\sim$~au).

\subsection{Geometrical modeling}
We used a geometrical approach to derive the basic characteristics of the NIR-emitting region. Given the exponential-like profile of PIONIER $V^2$ measurements, we considered a Lorentzian-like brightness distribution to represent the NIR emission, independently in {\itshape H} and {\itshape K}~bands. This centrally peaked brightness profile has broader tails than the usual Gaussian profile. It can be used to estimate a characteristic size for a spatially resolved emitting region that presents a gradually decreasing brightness profile with smooth outer limits. This is relevant for HD~139614, given its expected spatially extended NIR-emitting region. The squared visibility of a Lorentzian-like circumstellar emission can be written as
\begin{equation}
V^2_{\rm circ}(u,v)=\left|\exp\left(-2\pi \frac{r_{\scriptscriptstyle 50\%}}{\sqrt{3}}\frac{B_{\rm eff}(i,PA)}{\lambda}\right)\right|^2,
\end{equation}
where $r_{\scriptscriptstyle 50\%}$ is defined as the angular radius of half-integrated flux (containing 50\% of the total flux), and $B_{\rm eff}(i,PA)$ is the effective baseline \citep[see, e.g.,][]{2014A&A...561A..26M} with $i$ and $PA$ the inclination and position angle of the circumstellar component. 
Then, the total visibility $V^2_{\rm tot}$ is
{\small $V^2_{\rm tot}(u,v)=|f_*(\lambda)V_*(u,v)+(1-f_*(\lambda))V_{\rm circ}(u,v)|^2$}, with $V_*=1$ the visibility of the unresolved star. Here, $f_*(\lambda)$ is the star-to-total flux ratio that is estimated at each wavelength using the Kurucz spectrum of Sect.~2.5 for the star and a single-temperature blackbody for the circumstellar contribution. 
To limit the number of free parameters, we only consider two STFRs $f_{*,1300}$ and $f_{*,1500}$, calculated with a 1300~K and a 1500~K blackbody emission (see Sect.~3.1). The free parameters of the model are $r_{\scriptscriptstyle 50\%}$ (in mas), $i$, and $PA$.
\subsubsection{PIONIER}
\begin{figure}[t]
 \centering
 \includegraphics[scale=0.4]{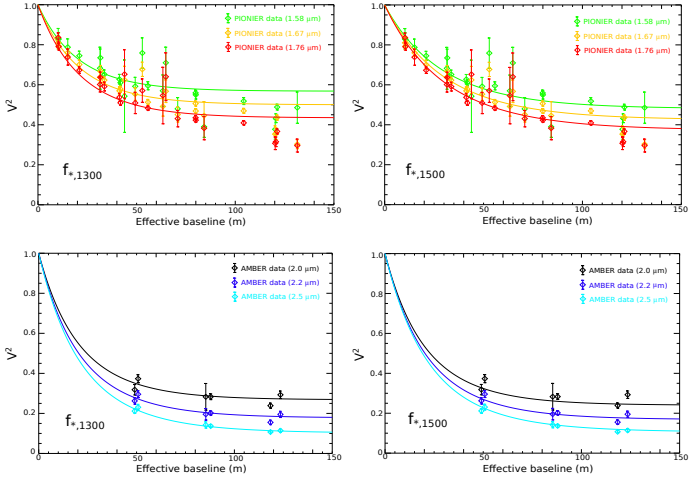}
 \caption{{\footnotesize Best-fit Lorentzian model (solid line) for PIONIER and AMBER overplotted on their measured $V^2$, as a function of the effective baseline (see Eq.1). For clarity, we only plot the AMBER $V^2$ at $\lambda=2.0$~$\mu$m, 2.2~$\mu$m, and 2.5~$\mu$m.}} 
 \label{fig:lorentzmodels}
\end{figure}

Considering the complete set of $V^2$ data, we computed a grid of models by scanning the parameters space in 70 steps and, from the $\chi^2$ calculated from the measured and modeled $V^2$, derived the Bayesian probability ($\exp \left[-\chi^2/2 \right]$) for each of the parameters. These marginal probability distributions are the projection of $\exp \left[-\chi^2/2 \right]$ along the three dimensions of the parameters space. The range of explored values is [1--20]~mas (0.1--2.8~au at 140~pc) for $r_{\scriptscriptstyle 50\%}$, [0$^{\circ}$--70$^{\circ}$] for $i$, and [0$^{\circ}$--175$^{\circ}$] for $PA$. It appears that low $i$ values ($\lesssim 35^{\circ}$) are favored, while no clear constraint is obtained on $PA$. Indeed, with $i \lesssim 35^{\circ}$, all $PA$ values between $0^{\circ}$ and $150^{\circ}$ fit the data almost equally well ($\chi^2_{\rm red} \simeq 2.3-2.6$). Since our available PIONIER dataset does not seem to highlight a significant difference in position angle and inclination relative to the outer disk, we adopted the values derived by \citet{2014A&A...561A..26M} for the outer disk: $112\pm9^{\circ}$ and $20\pm2^{\circ}$ (with $1-\sigma$ uncertainties), respectively. We
again explored the same range of values for $r_{\scriptscriptstyle 50\%}$ for the two STFRs. The best-fit solution is represented by $r_{\scriptscriptstyle 50\%}=3.9\pm 0.1$~mas ($0.55\pm0.01$~au at 140~pc) for $f_{*,1300}$, and by $r_{\scriptscriptstyle 50\%}=2.7\pm 0.2$~mas ($0.4\pm0.02$~au at 140~pc) for $f_{*,1500}$. The 1-$\sigma$ uncertainties correspond to the 68\% confidence interval derived directly from the marginal probability distribution. Using the $f_{*,1500}$ ratio ($f_{*,1500} < f_{*,1300}$ in {\itshape H} band) leads to a better fit ($\chi^2_{\rm red}=2.6$) than the $f_{*,1300}$ ratio ($\chi^2_{\rm red}=6.6$). As shown in Fig.\ref{fig:lorentzmodels}, the $V^2$ decrease at low spatial frequencies is reproduced in both cases, while the modeled $V^2$ for $f_{*,1300}$ significantly overestimates the $V^2$ measurements at the highest spatial frequencies. This suggests a discrepancy between the STFR predicted by the asymptotic $V^2$ level and the one estimated from the 2MASS measurements and the Kurucz spectrum. Possible causes are 1) the uncertainties on the 2MASS measurements \citep[$\sim 5\%$][]{2006AJ....131.1163S} and on the stellar parameters of the Kurucz model; 2) a change in the STFR between the time of the 2MASS and PIONIER observations, which however, appears unlikely given the absence of significant visible or MIR variability \citep{1998A&A...329..131M,2012ApJS..201...11K} and the face-on orientation of the disk \citep[see][ and references therein]{2014A&A...561A..26M}; and 3) the degradation of $V^2$ measurements at long baselines, where the coherent flux is lower. HD~139614 has a $H$ mag=7.3 that is close to the PIONIER sensitivity limit \citep[$H\sim 7$, ][]{2011A&A...535A..67L}.
Finally, no fully resolved emission was needed to reproduce the steep $V^2$ decrease at low spatial frequencies. 
\subsubsection{AMBER}
 Considering the complete set of AMBER dispersed $V^2$ data, we followed the same procedure as for PIONIER (same parameter space and same derivation of the formal errors). It quickly appeared that our AMBER interferometric dataset is too sparse in UV coverage (one baseline triplet covering only projected baselines lengths longer than 50~m) to constrain the inclination and the position angle of the {\itshape K}-band emitting region. Therefore, assuming again that the inner component is coplanar with the outer disk, we explored a broad range of $r_{\scriptscriptstyle 50\%}$ values (1--20~mas) and found a best-fit solution represented by $r_{\scriptscriptstyle 50\%}=4.3\pm 0.1$~mas ($0.60 \pm 0.02$~au at 140~pc) for $f_{*,1300}$, and by $r_{\scriptscriptstyle 50\%}=4.2\pm 0.1$~mas ($0.58 \pm 0.02$~au at 140~pc) for $f_{*,1500}$. As shown in Fig.\ref{fig:lorentzmodels}, both solutions agree well with the measured $V^2$ ($\chi^2_{\rm red}\sim1.6$), since $f_{*,1300}(\lambda) \simeq f_{*,1500}(\lambda)$ across the {\itshape K}-band. \\ 
 
\noindent Our new NIR data spatially resolved the innermost region of HD~139614 and suggest that hot dust material is located around the expected dust sublimation radius of micron-sized silicate dust ($\simeq 0.2$~au). We do not rule out the presence of refractory dust material within this radius. Moreover, the inner dust component probably extends out to 1~au or 2~au. Indeed, our derived $r_{\scriptscriptstyle 50\%}$ values imply that, for instance, 80\% of the flux in such Lorentzian profiles would then be contained within $\sim 1.1$~au in the {\itshape H} band, and $\sim 1.7$~au in the {\itshape K} band. Considering the uncertainty on the stellar distance ($\pm 27$~pc), the $r_{\scriptscriptstyle 50\%}$ values would vary from 0.35~au to 0.45~au in the {\itshape H} band and from 0.5~au to 0.7~au in  the {\itshape K} band. These values are still consistent with our conclusions. The slight difference in size estimate between the {\itshape H} and {\itshape K} bands suggests a disk chromaticity that is probably related to a temperature gradient. Finally, the data do not allow us to constrain a difference in inclination and position angle between the inner and outer components and are consistent with a coplanarity. 
 

\section{Radiative transfer modeling}
\subsection{Disk model}
Based on \citet{2014A&A...561A..26M}, we consider a two-component model composed of an inner and an outer dust component spatially separated by a dust-depleted region. Although we first assume an empty gap when performing the model's grid computation and $\chi^2$ minimization, an upper limit on the dust mass in the gap will then be estimated. We mention that the results of \citet{2014A&A...561A..26M}, which showed an incompatibility of the IR SED and the mid-IR visibilities with a continuous disk structure, confirmed previous results based on SED modeling. Notably, the possibility of a partially self-shadowed continuous disk was explored by \citet{2003A&A...398..607D}. Using a passive disk model with a puffed-up inner rim, they could not reproduce both the NIR excess and the rising MIR spectrum of HD\,139614 (see their Fig.4). The artificial increase in the inner rim scale height, which is required to match the NIR excess, implied a strong shadowing of the outer disk. This led to decreasing and too low MIR emission. We performed radiative transfer modeling tests with a continuous disk, including a puffed-up and optically thick inner rim to induce self-shadowing over the first few au. However, this model systematically produced too much flux in the 5--8~$\mu$m region, and decreasing and too weak emission at $\lambda > 8$~$\mu$m. This implied MIR visibilities without sine-like modulation, in contrast with the MIDI data. Moreover, the puffed-up inner rim systematically induced a too spatially confined NIR-emitting region that is incompatible with the NIR interferometric data.    
\subsubsection{RADMC3D}
We used the radiative transfer code RADMC3D to produce disk images and SED \citep{2004A&A...421.1075D}. Its robustness and accuracy were validated through benchmark studies \citep[e.g., ][]{2009A&A...498..967P}. This code can compute the dust temperature distribution using the Monte-Carlo method of \citet{2001ApJ...554..615B} with improvements like the continuous absorption method of \citet{1999A&A...345..211L}. We considered an axially symmetric two-dimensional disk in a polar coordinate system ($r$, $\theta$) with a logarithmic grid spacing in $r$ and $\theta$. An additional grid refinement in $r$ was applied to the inner edge of both components to ensure that the first grid cell is optically thin. The radiation field and temperature structure computed by the Monte Carlo runs were used to produce SEDs and images by integrating the radiative transfer equation along rays (ray-tracing method). Isotropic scattering was included in the modeling.  While the thermal source function is known from a first Monte Carlo run, the scattering source function is computed at each wavelength through an additional Monte Carlo run prior the ray-tracing.\subsubsection{Disk structure} Each dust component is represented by a parameterized model of passive disk with a mass $M_{\rm dust}$ and inner and outer radii, $r_{\rm in}$ and $r_{\rm out}$. Assuming it is similar to the gas density distribution in hydrostatic equilibrium, the dust density distribution is given by
\begin{equation}
\rho(r,z)=\frac{\Sigma(r)}{H(r)\sqrt{2\pi}}exp\left[-\frac{1}{2}\left(\frac{z}{H(r)}\right)^2\right],
\label{eq:dustdensity}
\end{equation}
where $z \simeq (\pi/2-\theta)r$ is the vertical distance from the midplane, in the case of a geometrically thin disk ($z >> r$).
The dust surface density $\Sigma(r)$ and scale height $H(r)$ are parameterized as
$\Sigma(r) = \Sigma_{\rm out}\left(r/r_{\rm out}\right)^p$ and $H(r) = H_{\rm out}\left(r/r_{\rm out}\right)^{1+\beta}$, where
$\beta$ is the flaring exponent, $\Sigma_{\rm out}$ and $H_{\rm out}$ are the dust surface density and scale height at $r_{\rm out}$. To enable a smooth decrease in density after $r_{\rm out}$, we apply another $p$ exponent to the surface density profile. We also include the possibility of rounding off or "puffing up" the inner rim of each component. For that, we artificially reduce or increase the dust scale height $H(r)$ to reach a chosen value $\hat{H}_{ \rm in}$ at $r_{\rm in}$. The width $\epsilon r_{\rm in}$ over which this reduction or increase is done -- from no change at $r_1 = (1+\epsilon)r_{\rm in}$ to the dust scale height $\hat{H}_{ \rm in}$ -- sets the ``sharpness" of the rim. Following \citet{2007A&A...471..173R}, the modified dust scale height $\hat{H}(r)$, between $r_{\rm in}$ and $r_1$, writes as
\begin{equation}
\hat{H}(r)=\left(1-\frac{r-r_{\rm in}}{r_1-r_{\rm in}}\right)\hat{H}_{\rm in}+\left(\frac{r-r_{\rm in}}{r_1-r_{\rm in}}\right)H(r).
\end{equation}
\subsubsection{Dust properties} The dust grain properties are described by their optical constants taken from the Jena database\footnote{Available at http://www.astro.uni-jena.de/Laboratory/OCDB/}. The mass absorption and scattering coefficients (in cm$^{2}$.g$^{-1}$), $\kappa_{\rm abs}(\lambda),$ and $\kappa_{\rm sca}(\lambda)$ are then computed using the Mie theory. The spherical shape approximation is usually safe for amorphous or featureless dust species for which the extinction properties are much less sensitive to grain shape effects than to crystalline material \citep[e.g., ][]{2005A&A...437..189V}. Assuming a grain size distribution $n(a)\propto a^{-3.5}$ \citep{1977ApJ...217..425M} with minimum $a_{\rm min}$ and maximum $a_{\rm max}$ grain sizes, the size-averaged mass absorption/scattering coefficient is obtained by adding the mass absorption/scattering coefficients of each grain size times their mass fraction. Then, the global $\kappa_{\rm abs}(\lambda)$ and $\kappa_{\rm sca}(\lambda)$ are derived by adding the size-averaged mass absorption/scattering coefficients of each dust species times their abundance to form a single "composite dust grain" that is the mix of the constituents. This averaged approach has already been justified in other disk studies \citep{2011A&A...531A..93M} and implicitly assumes a thermal coupling between the grains. This is expected since dust grains in disks are likely to be in the form of mixed aggregates in thermal contact. Each disk component has a homogeneous composition in the radial and vertical directions. No dust settling or radial segregation is considered here.  
\subsection{Modeling approach}
We split our modeling approach into several steps, considering the mutual radiative influence between the different disk components and the associated dataset. For the star emission, we use the synthetic spectrum detailed in Sect.~2.5 and Table~\ref{tab:star}. 
\subsubsection{Inner disk}
We address first the inner disk to reproduce the NIR SED and the PIONIER and AMBER $V^2$. Following Sect. 3.2, we set the inner disk's inclination and position angle to $i_{\rm disk}=20^{\circ}$ and $PA_{\rm disk}=112^{\circ}$. The inner rim is modeled as a vertical wall.\\
For the radial structure, $r_{\rm in}$ is first set to 0.2~au \citep[][]{2014A&A...561A..26M}. We vary the surface density profile exponent $p$ to explore different dust radial distributions. We also prevent $\Sigma(r)$ from decreasing too abruptly after $r_{\rm out}$ by setting $p=-10$. Greater $p$ values (e.g., $\sim -5$) would induce too smooth a decrease, keeping $r_{\rm out}$ from representing the inner disk's size.\\   
For the vertical structure, the inner disk scale height at $r_{\rm out}$ is assigned a broad range of values , namely $H_{\rm out}/r_{\rm out}=[0.03,0.05,0.1.0.2,0.3]$. This encompasses the dust scale height values that are typically inferred or considered (0.1--0.15) for disks \citep[e.g., ][]{2007ApJ...664L.107B,2010A&A...511A..75B}.\\
\begin{table}[t]
\caption{Description of the dust setups}
 \centering
 \begin{tabular}{ccccc}
 \hline
 \hline
 {\footnotesize Dust setup}&{\footnotesize Dust species}&{\footnotesize $a_{\rm min}$~$(\mu$m)}&{\footnotesize $a_{\rm max}$~$(\mu$m)}&{\footnotesize Abundance}\vspace{1mm}\\
 
 \multirow{2}{*}{{\scriptsize 1}}&{\scriptsize Olivine}&{\scriptsize 0.1}&{\scriptsize 20}&{\scriptsize 80\%}\\
 &{\scriptsize Graphite}&{\scriptsize 0.05}&{\scriptsize 0.2}&{\scriptsize 20\%}\\
 \multirow{2}{*}{{\scriptsize 2}}&{\scriptsize Olivine}&{\scriptsize 5}&{\scriptsize 20}&{\scriptsize 80\%}\\
 &{\scriptsize Graphite}&{\scriptsize 0.05}&{\scriptsize 0.2}&{\scriptsize 20\%}\\
 {\scriptsize 3}&{\scriptsize Olivine}&{\scriptsize 0.1}&{\scriptsize 20}&{\scriptsize 100\%}\\
 {\scriptsize 4}&{\scriptsize Olivine}&{\scriptsize 5}&{\scriptsize 20}&{\scriptsize 100\%}\\ 
 {\scriptsize 5}&{\scriptsize Graphite}&{\scriptsize 0.05}&{\scriptsize 0.2}&{\scriptsize 100\%}\\
 {\scriptsize 6}&{\scriptsize Olivine}&{\scriptsize 0.1}&{\scriptsize 3000}&{\scriptsize 100\%}\\
 \hline
 \end{tabular}
 \tablefoot{$a_{\rm min}$ and $a_{\rm max}$ are the minimum and maximum grain sizes . The abundance is in mass. The dust setup `6' is only used for the outer disk.}
 \label{tab:dustsetup}
 \end{table} 
\noindent We consider a dust mix of silicates and graphite. Including graphite is justified by its being a major constituent of the interstellar grains \citep{2003ARA&A..41..241D} and by recent disks modeling that shows the importance of featureless and refractory species like graphite to explain the observations \citep{2002A&A...392.1039M,Carmona2014}. We considered three graphite mass fractions (see Table~\ref{tab:dustsetup}) to evaluate its impact on the modeling. We assumed the same graphite composition as in \citet{2008A&A...478..779S} and considered grains with $a_{\rm min}=0.05$~$\mu$m and $a_{\rm max}=0.2$~$\mu$m to maximize the IR opacity. Since graphite is a featureless species, this choice is not critical. For the silicate content, we assumed a pure iron-free olivine composition \citep[see ][]{2010ApJ...721..431J} with $a_{\rm min}=0.1$~$\mu$m and 5~$\mu$m to keep or suppress the silicate feature. We fixed $a_{\rm max}=20$~$\mu$m since the optical and NIR opacity contribution from mm grains is negligible.\\
The inner disk free parameters are $p$, $r_{\rm out}$, and $M_{\rm dust}$. The last is a lower limit since we only include $\mu$m-sized and sub-$\mu$m-sized grains. Then, the modeling steps are
\begin{itemize} 
\item computing a grid of $10\times10\times10$ models on $M_{\rm dust}$, $p$, $r_{\rm out}$ for each model setup (combination of a dust setup and a $H_{\rm out}/r_{\rm out}$ value; see Table~\ref{tab:freeparams}). We then calculate a reduced $\chi^2_{\rm r}$ separately for each dataset: the NIR SED (5 points from 1.25~$\mu$m to 4.5~$\mu$m), the dispersed PIONIER $V^2$, and dispersed AMBER $V^2$. The synthetic $V^2$ are computed from the RADMC3D images multiplied by a 2D Gaussian with a FWHM equal to the instrument FOV. The best-fit model of each model setup is found by minimizing the sum of the reduced $\chi^2_r$ values. 
\item  for each model setup, computing a grid of $8\times8\times8$ models around the global minimum $\chi^2_r$ and calculating the marginal probability distribution ($\exp[-\chi^2/2]$) to determine the best-fit value and 1-$\sigma$ uncertainty (68\% confidence interval) on $M_{\rm dust}$, $p$, and $r_{\rm out}$. The best-fit model with the lowest reduced $\chi^2_r$ value is chosen as the global best-fit solution.
\item Once the global best-fit solution is found, slightly varying $r_{\rm in}$ around its initial value ($0.2$~au) to further improve the fit to the NIR visibilities at high spatial frequencies.  
\end{itemize}
\subsubsection{Outer disk}
We then address the outer disk to find a solution that is consistent with the Herschel/PACS data and the sub-mm SED. Based on the model of \citet{2014A&A...561A..26M}, we set $r_{\rm in}=5.6$~au and $r_{\rm out}=150$~au and assume a pure-olivine composition for the silicate content, with a grain size distribution from 0.1~$\mu$m to 3~mm to account for the weak silicate emission feature in the {\itshape N} band and the low sub-mm spectral index \citep[][]{1996MNRAS.279..915S}. We set the surface density profile exponent to $p=-1$, as is typically observed in the outer regions of disks \citep[e.g., ][]{2007ApJ...659.705A,2010ApJ...723.1241A}. Then, two typical flaring profiles for irradiated disks are explored, i.e., $\beta=1/7$ and $\beta=2/7$ \citep[e.g., ][]{1997ApJ...490..368C}, and the dust scale height $H_{\rm out}/r_{\rm out}$ is varied between 0.1 and 0.15, which is typical of disks \citep[e.g.,][]{2010A&A...511A..75B}. We also vary the graphite abundance (from 0 to 20\%) and the dust mass between 10$^{-6}$~M$_{\odot}$ and 10$^{-4}$~M$_{\odot}$.
\begin{figure}[t]
 \centering
 \includegraphics[scale=0.5]{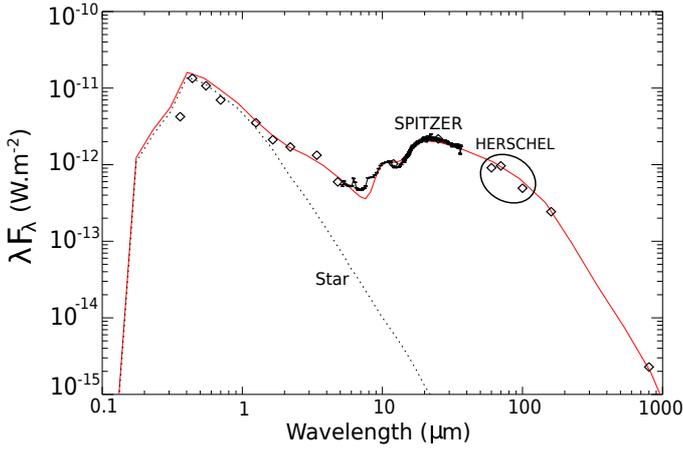}
  \caption{{\footnotesize {\itshape Top left}: observed SED (black diamonds) and SED of the best-fit RADMC3D model (red line). The {\itshape SPITZER}/IRS spectrum is indicated with a black line, and the stellar contribution as a dotted line.}}
 \label{fig:radmcsed}
\end{figure}
\subsubsection{Outer disk inner edge} 
We then focus on the inner edge to reproduce the MIDI data and improve the fit to the MIR SED. Both trace the geometry and flux fraction intercepted by the outer disk's inner edge, hence its scale height and radial position. Since the SED measurements at 4.8~$\mu$m, 12~$\mu$m, and 25~$\mu$m are consistent with the {\itshape Spitzer} spectrum, modeling and fitting them will be sufficient and lead to a decent fit of the {\itshape Spitzer} spectrum. Reproducing the full {\itshape Spitzer} spectrum
in detail is beyond the scope of this paper.
Using the global best-fit solution of the inner disk, we compute a grid of $8\times8\times8$ models on the outer disk inner radius $r_{\rm in}$, the modified scale height $\hat{H}_{\rm in}$ at $r_{\rm in}$, and the rounding-off parameter $\epsilon$ (see Table~\ref{tab:freeparams}). We then calculate a reduced $\chi^2_{\rm r}$ separately for each dataset, namely the dispersed MIDI visibilities (47 data points between 8~$\mu$m and 13~$\mu$m) and the MIR SED (at 4.8~$\mu$m, 12~$\mu$m, and 25~$\mu$m).
The best-fit model is found by minimizing the sum of the reduced $\chi^2_{\rm r}$ values. The parameters' best-fit values and 1-$\sigma$ uncertainties are computed in the same way as for the inner disk.

\begin{table}[t]
\caption{Model setups and range of free parameters values explored.}
 \centering
 \begin{tabular}{ccccc}
 \hline
 \hline
 \multicolumn{5}{c}{{\footnotesize Inner disk}}\\
 \hline
 \multicolumn{2}{c}{{\footnotesize Model setups}}&\multicolumn{3}{c}{{\footnotesize Free parameters}}\\
 \hline  
 {\footnotesize Dust setup}&{\footnotesize $H_{\rm out}/r_{\rm out}$}&{\footnotesize $r_{\rm out}$ (au)}&{\footnotesize $M_{\rm dust}$ (M$_{\odot}$)}&{\footnotesize $p$}\\
 {\scriptsize 1}&{\scriptsize [0.03--0.3]}&{\scriptsize [1.0--3.0]}&{\scriptsize [$10^{-10.0}\,$--$\,10^{-7.5}$]}&{\scriptsize [(-1.5)--3.0]}\\
 {\scriptsize 2}&{\scriptsize [0.03--0.3]}&{\scriptsize [1.0--3.0]}&{\scriptsize [$10^{-9.5}\,$--$\,10^{-7.0}$]}&{\scriptsize [(-1.5)--3.0]}\\
 {\scriptsize 3}&{\scriptsize [0.03--0.3]}&{\scriptsize [1.0--3.0]}&{\scriptsize [$10^{-8.0}\,$--$\,10^{-5.5}$]}&{\scriptsize [(-1.5)--3.0]}\\
 {\scriptsize 4}&{\scriptsize [0.03--0.3]}&{\scriptsize [0.5--2.5]}&{\scriptsize [$10^{-9.5}\,$--$\,10^{-6.0}$]}&{\scriptsize [(-1.0)--4.0]}\\
 {\scriptsize 5}&{\scriptsize [0.03--0.3]}&{\scriptsize [1.0--3.0]}&{\scriptsize [$10^{-11.0}\,$--$\,10^{-7.5}$]}&{\scriptsize [(-1.5)--3.0]}\\
 \hline
 \multicolumn{5}{c}{{\footnotesize Outer disk's inner edge}}\\
  \hline
 \multicolumn{2}{c}{{\footnotesize Model setups}}&\multicolumn{3}{c}{{\footnotesize Free parameters}}\\
 \hline  
 \multicolumn{2}{c}{{\footnotesize Dust setup}}&{\footnotesize $\hat{H}_{\rm in}$}&$\epsilon$&{\footnotesize $r_{\rm in}$ (au)}\\
 \multicolumn{2}{c}{{\scriptsize 6}}&{\scriptsize [0.01--0.07]}&{\scriptsize [0.1--0.6]}&{\scriptsize [5.4--6.1]}\\
 \hline
 \end{tabular}
 \label{tab:freeparams}
 \end{table} 
\subsection{Results}
\begin{figure*}[t]
 \centering
 \resizebox{\hsize}{!}{\includegraphics[width=80mm,height=50mm]{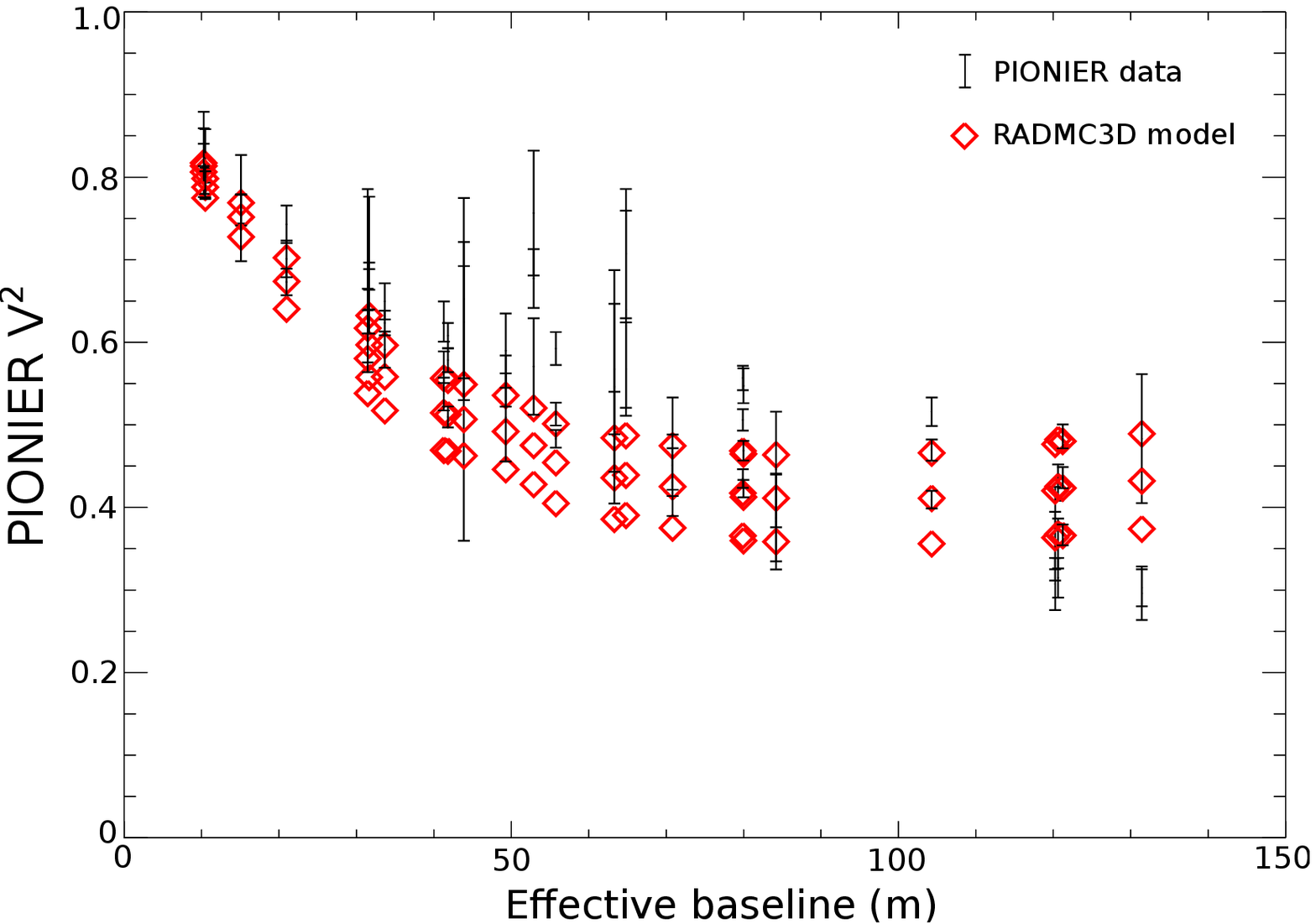}
 \includegraphics[width=80mm,height=50mm]{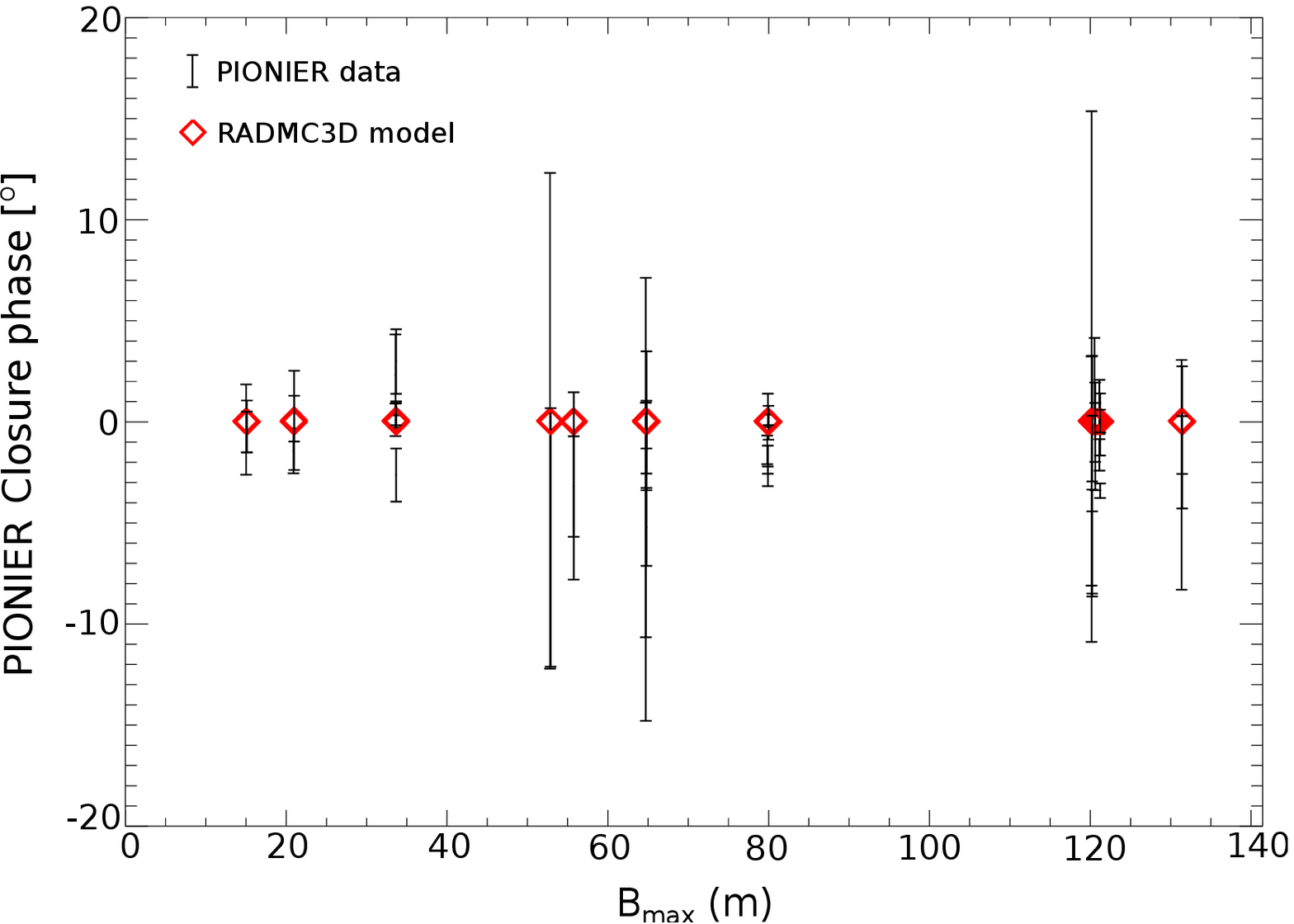}}\\[1.5mm] 
 \resizebox{\hsize}{!}{\includegraphics[width=90mm,height=55mm]{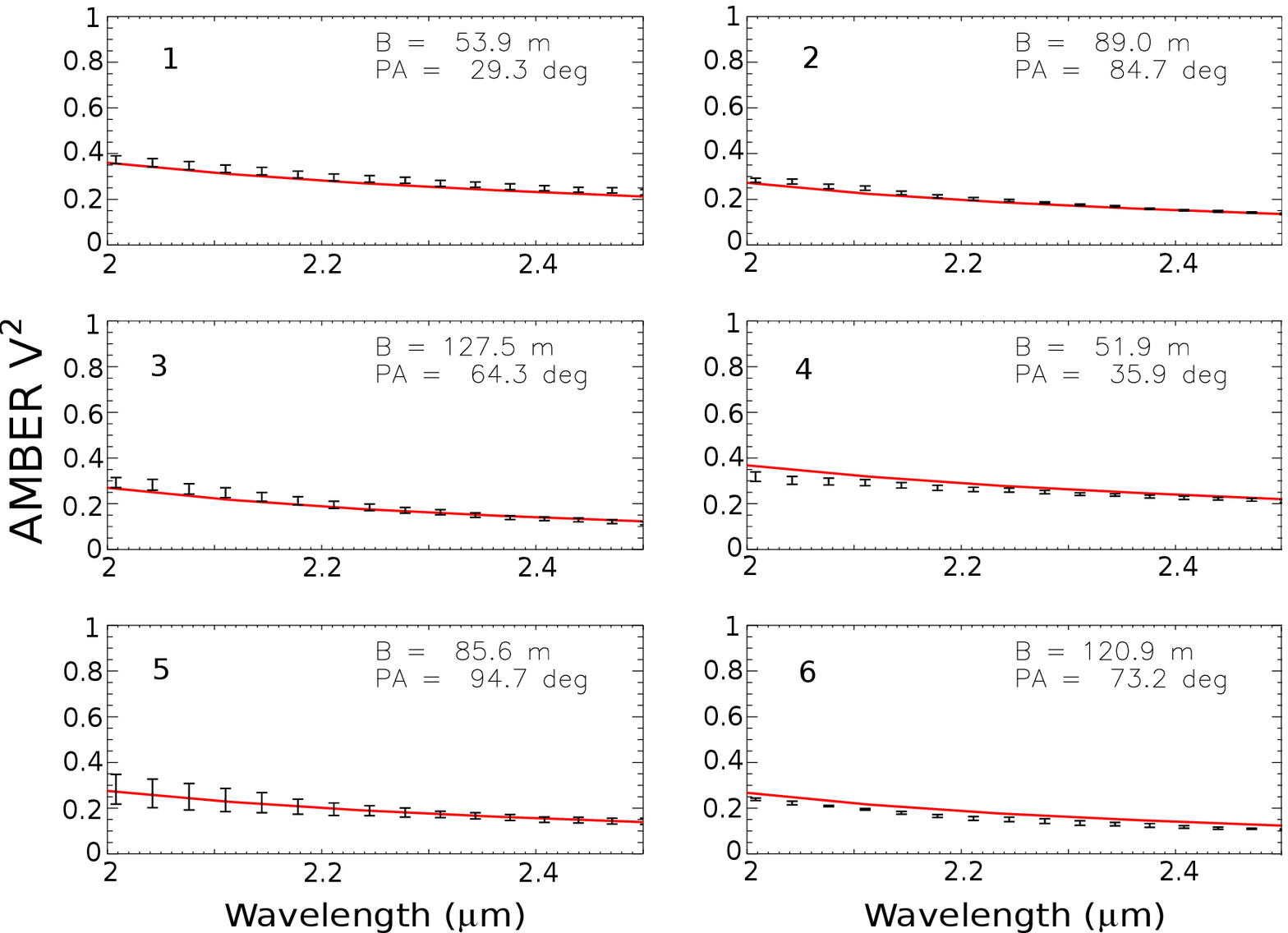}\hfill
 \includegraphics[width=90mm,height=55mm]{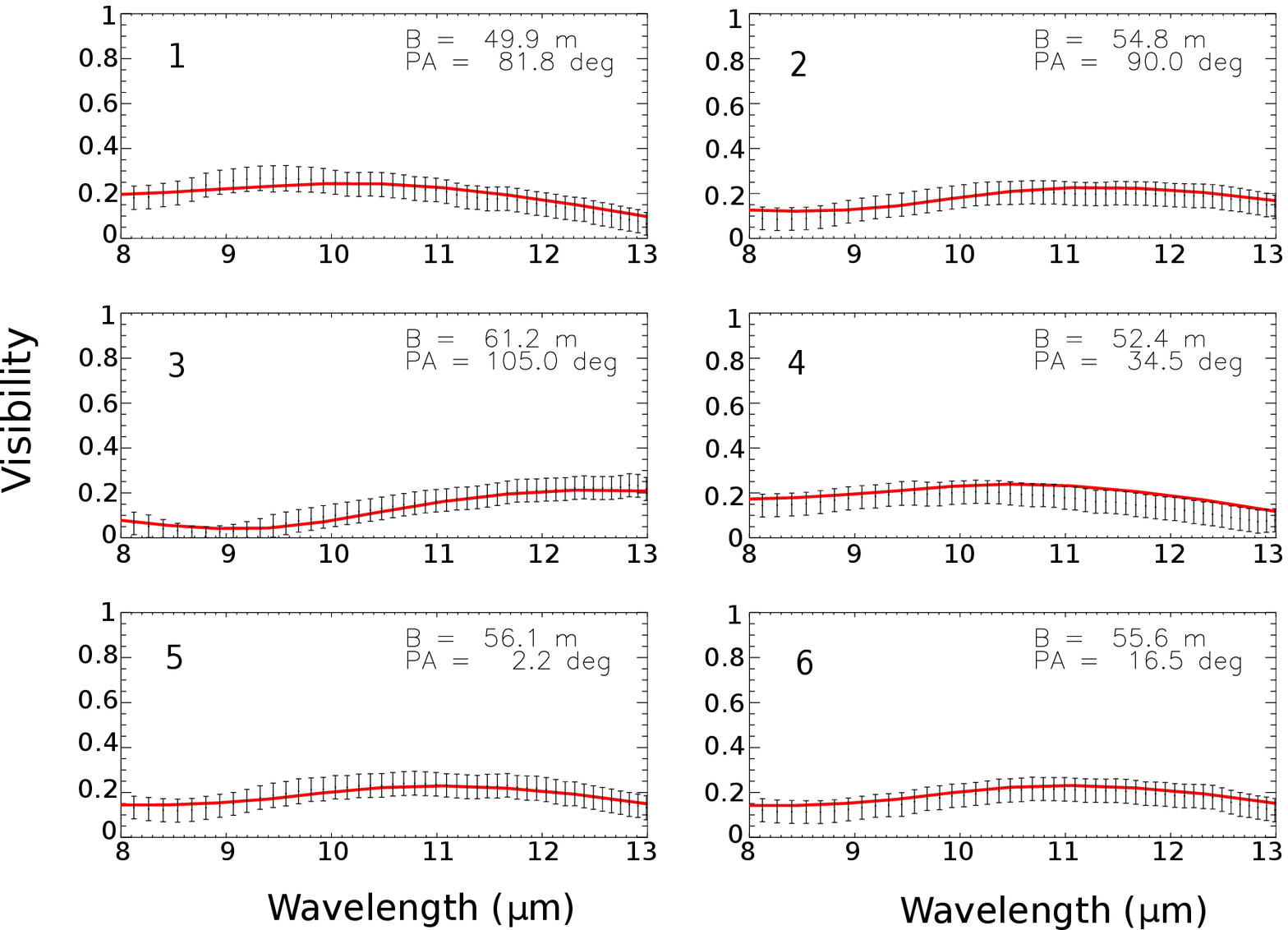} }
  \caption{{\footnotesize {\itshape Top left}: Best-fit {\itshape H}-band $V^2$ from the radiative transfer (red diamonds) and PIONIER $V^2$. {\itshape Top right}: Same for the PIONIER closure phases that were only used to check for consistency (B$_{\rm max}$ is the longest baseline of the triplet). {\itshape Bottom left}: Best-fit {\itshape K}-band $V^2$ from the radiative transfer (red line) and AMBER $V^2$ (black). {\itshape Bottom right}: same for the modeled {\itshape N}-band visibilities and the MIDI data (detailed in \citet{2014A&A...561A..26M}). }}
 \label{fig:radmcmodel}
\end{figure*}

\subsubsection{A tenuous and extended inner disk}
\begin{figure*}[t]
 \centering
 \includegraphics[width=130mm,height=45mm]{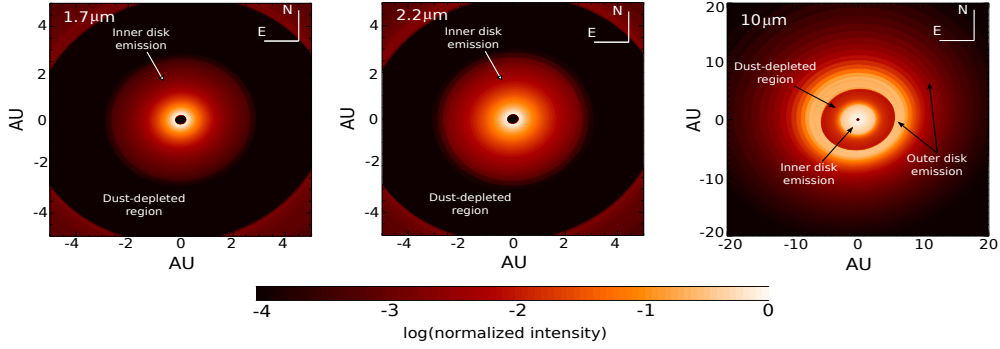}
  \caption{{\footnotesize Synthetic images of our best-fit RADMC3D model at $\lambda=1.7$~$\mu$m (PIONIER), $\lambda=2.2$~$\mu$m (AMBER), and $\lambda=10$~$\mu$m (MIDI). For each image, the intensity is normalized to one at maximum and represented on logarithmic scale (central star is removed).}}
 \label{fig:imagemodel}
\end{figure*}

\begin{figure}[t]
 \centering
 \includegraphics[width=70mm,height=44mm]{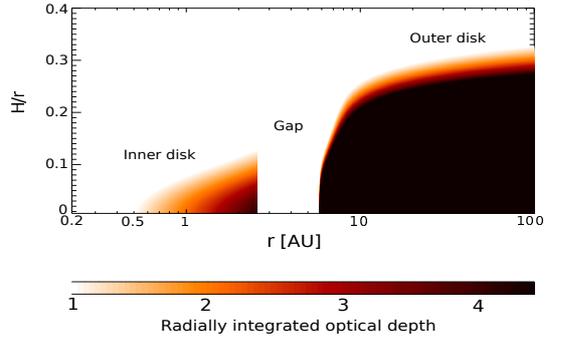}
  \caption{{\footnotesize Radially integrated optical depth profile, at $\lambda=0.55$~$\mu$m (stellar emission peak), of the best-fit model. For clarity, we articially cut the integrated optical depth profile of the inner disk at $r_{\rm out}=2.55$~au.}}
 \label{fig:taudisk}
\end{figure}

\begin{table}[t]
\caption{Best-fit model setups for the inner disk.}
 \centering
 \begin{tabular}{cccccc}
 \hline
 \hline
 \multicolumn{2}{c}{{\footnotesize Model setup}}&\multicolumn{3}{c}{{\footnotesize Free parameters}}\\
 \hline  
 {\footnotesize Dust setup}&{\footnotesize $H_{\rm out}/r_{\rm out}$}&{\footnotesize $r_{\rm out}$ (au)}&{\footnotesize $M_{\rm dust}$ (M$_{\odot}$)}&{\footnotesize $p$}&{\footnotesize $\chi^2_{\rm r}$}\\[1mm]
 \multirow{2}{*}{{\scriptsize 2}}&{\scriptsize 0.05}&{\scriptsize $1.89^{+0.14}_{0.13}$}&{\scriptsize $2.8^{+0.8}_{-0.8}\times10^{-10}$}&{\scriptsize $1.6^{+0.3}_{-0.3}$}&3.9\\[1mm]
 &{\scriptsize 0.1}&{\scriptsize $2.55^{+0.13}_{-0.13}$}&{\scriptsize $8.8^{+4.2}_{-3.7}\times10^{-11}$}&{\scriptsize $0.6^{+0.3}_{-0.3}$}&4.0\\[1mm]
 \multirow{2}{*}{{\scriptsize 5}}&{\scriptsize 0.03}&{\scriptsize $1.10^{+0.11}_{0.11}$}&{\scriptsize $9.5^{+3.1}_{-3.1}\times10^{-11}$}&{\scriptsize $3.4^{+0.3}_{-0.3}$}&4.4\\[1mm]
 &{\scriptsize 0.05}&{\scriptsize $2.20^{+0.10}_{-0.10}$}&{\scriptsize $4.2^{+1.2}_{-1.2}\times10^{-11}$}&{\scriptsize $1.1^{+0.3}_{-0.3}$}&4.0\\[1mm]
 \hline
 \end{tabular}
 \tablefoot{These best-fit model setups (with $r_{\rm in}$ set to 0.2~au) fit the NIR data ($\chi^2_{\rm r}<5$) equally. The quoted uncertainties are 1-$\sigma$ (68\% confidence interval). The MIR data were not used here.}
 \label{tab:inner}
 \end{table}
\noindent Table~\ref{tab:inner} shows the results of the inner disk modeling. Several models with different $r_{\rm out}$ values for the inner disk fit the NIR data equally well. Since our MIDI data partly resolve the inner disk \citep{2014A&A...561A..26M}, we can use them to break out the degeneracy in $r_{\rm out}$ and identify the global best-fit solution. For each inner disk solution, we computed a grid of $8\times8\times8$ models on the outer disk's inner edge parameters, and kept the model giving the lowest reduced $\chi^2_{\rm r}$ on the MIR SED and visibilities.
\paragraph{Size and location}
Only the inner disk models with $r_{\rm out} > 2$~au are compatible with the MIR visibilities, especially from 8 to 9.5~$\mu$m. We constrained the inner disk's outer radius to $r_{\rm out}=2.55^{+0.13}_{-0.13}$~au. With an uncertainty of ($\pm 27$~pc) on the stellar distance, $r_{\rm out}$ varies from 2.0 to 2.9~au, which is still consistent with our conclusions. This small a variation in $r_{\rm out}$ would not significantly modify the conditions of irradiation of the inner disk and thus its temperature and brightness profile. 
From the best-fit solution, we varied $r_{\rm in}$ by steps of 0.05~au around $r_{\rm in}=0.2$~au to optimize the fit to the PIONIER and AMBER $V^2$. The best agreement is nevertheless found for $r_{\rm in}=0.2$~au. The $r_{\rm in}<0.2$~au values slightly increased the amount of unresolved inner disk emission and induced too high a {\itshape K}-band $V^2$ at low spatial frequencies. The $r_{\rm in}>0.2$~au values induced too prominent a lobe in the {\itshape H}-band $V^2$ profile at high spatial frequencies (see Appendix).  
 As shown in Figs.\ref{fig:radmcsed} and \ref{fig:radmcmodel}, our model agrees well with the NIR SED and the PIONIER and AMBER $V^2$. Nevertheless, the best-fit {\itshape H} band $V^2$ are still slightly too high at high spatial frequencies, which probably suggests that the inner rim has a smoother shape than the assumed vertical wall. The modeled closure phases in the {\itshape H} band appears consistent with the measured ones. 
Figure~\ref{fig:imagemodel} shows the best-fit model images at $\lambda=1.6$~$\mu$m and $\lambda=2$~$\mu$m. 
\paragraph{Dust composition}
Our modeling favored the dust setup "2" (80\% olivine + 20\% graphite, see Table ~\ref{tab:dustsetup}). 
%
%
This implies a temperature for the hottest composite grains of $T\simeq1660$~K at $r_{\rm in}=0.2$~au, which is on the warm side of sublimation temperatures for $\mu$m-sized silicates \citep{1994ApJ...421..615P}. We also tested the effect of computing the temperature distribution separately for the two dust species. This led to olivine grains that were too cold ($\sim 1150$~K) to be responsible for the NIR excess and to very hot graphite grains ($\sim 1900$~K). The latter would induce too much flux at $\lambda < 2$~$\mu$m and would disagree with our conclusions in Sect.~3.1.
Therefore small graphite grains (or any refractory, featureless and efficient absorber/emitter) in thermal contact with the silicate grains seems required to induce enough heating and emission in the inner disk's outer parts ($> 1$~au) and to produce a spatially extended NIR-emitting region. The solution with a 100\% graphite composition and $r_{\rm out}=2.20^{+0.10}_{-0.10}$~au is also compatible with the MIR visibilities. The dust compositions including smaller (sub-$\mu$m-sized) olivine grains (dust setups "1" and "3") induce high inner rim temperatures ($\sim 1550$~K), for which these small grains may sublimate, and overpredict the strength of the 10~$\mu$m feature in the {\itshape Spitzer} spectrum. This suggests that the inner disk does not contain a detectable amount of sub-$\mu$m-sized silicate grains and that the weak silicate feature originates in the outer disk. 
\paragraph{Surface density profile}
We constrained the surface density profile to $p=0.6^{+0.3}_{-0.3}$ (see Fig.\ref{fig:surfacedensity}). This positive profile suggests a radially increasing distribution of dust grains, and was required to create the extended NIR-emitting region predicted by the interferometric data. Such a radial distribution induces less shadowing from the inner rim and sufficient heating and re-emission from the outer parts of the inner disk. The radially integrated midplane optical depth (at $\lambda=0.55$~$\mu$m) reaches $\tau \simeq 1$ only at $r \simeq 0.6$~au, as shown in Fig.\ref{fig:taudisk}. All the models with $p<0$ induced too strong a flux contribution from the inner disk rim in the SED at $\lambda \lesssim 2$~$\mu$m and therefore a NIR-emitting region that is too spatially confined. This implied {\itshape H}-band and {\itshape K}-band $V^2$ that are too high at low spatial frequencies, and too low at high spatial frequencies since the STFR is underestimated (see Appendix).
\paragraph{Dust mass and surface density level}
We estimated a mass of $M_{\rm dust}=8.8^{+4.2}_{-3.7}\times10^{-11}$~M$_{\odot}$ in small dust grains, which dominate the optical/IR extinction efficiency, and set the NIR emission level. The NIR absorption efficiency is dominated by the small graphite grains with a mass absorption coefficient 100 times larger than for the $\mu$m-sized olivine grains. Therefore the total dust mass estimate largely depends on the graphite fraction, and a significant amount of silicate grains could be hidden in the inner disk. 
To estimate an upper limit on the dust mass, we extended the olivine grain size distribution up to 3000~$\mu$m (as in the outer disk) and kept the same absolute mass in graphite grains of our best-fit model. We then increased the mass in olivine grains (by steps of $10^{-10}$~M$_{\odot}$) until the modeled NIR emission (at 2.2~$\mu$m, 3.4~$\mu$m, and 4.8~$\mu$m) and/or NIR visibilities deviated by more than 3-$\sigma$ from the observed ones. As a result, we found an upper limit of about $6\times10^{-10}$~M$_{\odot}$.
Higher mass values implied a deviation of more than 3-$\sigma$ with respect to several AMBER $V^2$ measurements (all across the {\itshape K} band) and to the observed SED at 4.8~$\mu$m.
We show in Fig.\ref{fig:surfacedensity} the best-fit dust surface density and overplot its upper limit (dotted line). 
Relative to the outer disk, the inner disk seems strongly depleted by at least $\sim 10^3$. However, this dust depletion may be biased by a difference in dust composition between the inner disk that contains graphite grains and the outer disk that only contains olivine grains (see Table~\ref{tab:dustsetup}). 
However, the NIR mass absorption coefficient of the inner disk's dust mixture is larger, by a factor 5, than that of the outer disk. In the MIR, the two coefficients are equal. The difference in dust composition thus cannot account for the dust surface density ratio of $\sim 10^3$ between the two components.
\paragraph{Dust scale height}
A dust scale height of $H_{\rm out}\simeq 0.25$~au ($H_{\rm out}/r_{\rm out}=0.1$) at $r_{\rm out}=2.55$~au is favored. This translates to $H_{\rm in}\simeq 0.01$~au ($H_{\rm in}/r_{\rm in}=0.044$) at the inner rim. 
We recall that $H(r)$ is the height from the midplane at which the dust density has decreased by a factor ${\rm e}^{-0.5}$ (see Eq.\ref{eq:dustdensity}). Assuming a perfect dust-gas coupling, this would equal the pressure scale height of the gas in hydrostatic equilibrium. With a midplane dust temperature of 1660~K at $r_{\rm in}$, the gas pressure scale height would be smaller than the dust scale height ($H_{\rm in,gas}=0.006$~au, i.e. $H_{\rm in,gas}/r_{\rm in} \simeq 0.028$). If dust grains are coupled to the gas, this suggests that 1) gas is actually hotter (with a required temperature of about 5000~K), as expected in the inner disk regions from thermo-chemical modeling \citep[e.g., ][]{2009A&A...501..383W}, and/or that 2) gas is vertically supported by additional sources, such as magnetic forces arising from the inner disk accretion driven by Magneto-rotational turbulence \citep{2014ApJ...780...42T}. With this dust scale height, the inner disk does not cast a significant shadow on the inner rim of the outer disk. Indeed, while the total integrated optical depth (at $\lambda=0.55$~$\mu$m) in the inner disk midplane is $\tau \simeq 4$, it decreases quickly to $\tau < 1$, above the midplane, so that the outer disk's inner rim can intercept a large fraction of the stellar irradiation (see Fig.\ref{fig:taudisk}). The impact can be clearly seen in the emission increase at $\lambda \leq 8$~$\mu$m in the best-fit model SED (see Fig.\ref{fig:radmcsed}) and in the ring-like morphology of the outer disk's rim in the synthetic image at 10~$\mu$m (see Fig.\ref{fig:imagemodel}). 
\begin{table*}[t]
\caption{Parameters values of the global best-fit RADMC3D model.}
 \centering
 \begin{tabular}{cccccccccccc}
 \hline
 \hline
 &{\footnotesize Dust setup}&{\footnotesize $H_{\rm out}/r_{\rm out}$}&{\footnotesize $r_{\rm in}$ [au]}&{\footnotesize $r_{\rm out}$ [au]}&{\footnotesize $M_{\rm dust}$ [M$_{\odot}$]}&{\footnotesize $p$}&{\footnotesize $\beta$}&{\footnotesize $\hat{H}_{\rm in}/r_{\rm in}$}&{\footnotesize $\epsilon$}&{\footnotesize $i_{\rm disk}$ [$^{\circ}$]}&{\footnotesize $PA_{\rm disk}$ [$^{\circ}$]}\\[1mm]
 {\small Inner disk}& {\footnotesize 2}& {\footnotesize 0.1}& {\footnotesize 0.20\tablefootmark{m}}&{\footnotesize $2.55^{+0.13}_{-0.13}$}& {\footnotesize $8.9^{+3.4}_{-2.5}\times10^{-11}$}& {\footnotesize $0.6^{+0.3}_{-0.3}$}& {\footnotesize 2/7\tablefootmark{m}}& {\footnotesize N/A}\tablefootmark{*}& {\footnotesize 0\tablefootmark{f}}& {\footnotesize 20\tablefootmark{f}}& 112\tablefootmark{f}\\[1mm]
 {\small Outer disk}& {\footnotesize 6\tablefootmark{m}}& {\footnotesize 0.135\tablefootmark{m}}& {\footnotesize $5.7^{+0.3}_{-0.2}$}&{\footnotesize 150\tablefootmark{f}}& {\footnotesize $1.3\times10^{-4}$\tablefootmark{\:m}}& {\footnotesize $-1$\tablefootmark{f}}& {\footnotesize 1/7\tablefootmark{m}}& {\footnotesize $0.04^{+0.02}_{-0.02}$}& {\footnotesize $0.30^{+0.15}_{-0.05}$}& {\footnotesize  20\tablefootmark{f}}& {\footnotesize 112\tablefootmark{f}}  \\
 \hline
 \end{tabular}
 \tablefoot{The free parameters are indicated with their 1-sigma uncertainty (68\% confidence interval). $i_{\rm disk}$ and $PA_{\rm disk}$ denote the disk inclination and position angle. \tablefoottext{f}{Parameter value fixed during the search for the best-fit model.} \tablefoottext{m}{Parameter value obtained manually, i.e. without $\chi^2$ calculation and minimization (see Sect.~4.2.1 and 4.2.2).} \tablefoottext{*}{No artificial modification of the dust scale height was applied at $r_{\rm in}$ for the inner disk.}}
 \label{tab:global}
 \end{table*}  


\subsubsection{An au-sized gap}
As shown in Table~\ref{tab:global}, the outer disk starts at 5.7~au, and the gap, depleted in warm small grains, is au-sized ($\sim$ 3.2 au), which confirms the results of \citet{2014A&A...561A..26M}. Considering the stellar distance uncertainty ($\pm 27$~pc), the outer disk's inner radius varies between 4.7~au and 6.7~au. The corresponding temperature variation is less than 5~K in the disk midplane and $\sim 10$~K in the higher disk surface. The gap width would range from 2.7~au to 3.8~au, which is still in the au-sized range. Therefore, we do not expect significant changes in the emission profile of the outer disk's inner edge, hence no impact on the modeled MIR SED and visibilities. 

The best-fit {\itshape N}-band visibilities are shown in Fig.\ref{fig:radmcmodel}. Both the level and the shape (sine-like modulation) in the measured visibilities are reproduced by our model. Figure~\ref{fig:imagemodel} shows the best-fit model image at $\lambda=10$~$\mu$m (right panel), which shows the extended inner disk and the ring-like morphology of the outer disk's inner rim.
Then, we estimate an upper limit on the dust mass that could lie within the gap. Gaps opened by substellar companions can in principle filter dust grains partially decoupled from the gas \citep[e.g.,][]{2006MNRAS.373.1619R}. Therefore, the dust in the gap and the inner disk likely differs from the outer disk. 

For the gap, we thus assumed the same dust composition (size distribution from 5~$\mu$m to 3000~$\mu$m for olivine grains and graphite fraction of 3\%) and scale height profile as for the inner disk and considered the simplest case of a constant dust surface density profile ($p=0$). We then varied the amount of dust in the gap until the modeled MIR SED (at 4.8~$\mu$m, 12~$\mu$m and 25~$\mu$m) and/or the MIR visibilities start deviating by more than 3-$\sigma$ relative to the observations. As a result, we estimated an upper limit of $\sim 2.5\times10^{-10}$~M$_\odot$. Above this value, the modeled SED overestimated the observed SED at 4.8~$\mu$m
by more than 3-$\sigma$, while two of the modeled MIR visibilities became lower, by more than 1-$\sigma$, than the MIDI visibilities between 11.5 and 13~$\mu$m. As the gap is being filled in, the amplitude of the sine-like modulation also decreases in the modeled MIR visibilities. Interestingly, the mass estimate inside the gap is comparable to the upper limit found for the inner disk. We recall that other mass reservoirs could lie within the gap, such as cold mm-sized grains, pebbles, or minor bodies, which do not significantly contribute to the disk IR emission.

\subsubsection{An outer disk with a "smooth" inner edge}
A $1.3\times10^{-4}$~M$_{\odot}$ outer dust disk with a flaring index $\beta=1/7$ and a scale height $H_{\rm out}/r_{\rm out}=0.135$ at $r_{\rm out}$ consistently reproduces the HERSCHEL/PACS and sub-mm photometry measurements. A 100\% olivine dust composition with 0.1~$\mu{\rm m}<a<3$~mm was sufficient to account for the {\itshape SPITZER}/IRS measurements and its weak 10~$\mu$m silicate feature, along with the sub-mm spectral slope. Adding graphite in our model was not required. The best-fit broadband SED is presented in Fig.\ref{fig:radmcsed}.\\
Our MIDI data favored a slightly rounded shape for the outer disk's inner rim over a purely vertical wall. A reduction of the scale height of the dusty disk from $1.2 \times r_{\rm in}=7.4$~au to $r_{\rm in}=5.7$~au ($\epsilon=0.3$), leading to $H_{\rm in}/r_{\rm in}=0.04$ (instead of $H_{\rm in}/r_{\rm in}=0.085$), was needed to reproduce the modulation in the measured MIDI visibilities. A purely vertical wall induced too strong a modulation with the apparition of a pronounced lobe for the third visibility measurement. The rim shape could be explored further in a future study using hydrodynamical simulations to infer the mass of an hypothetical substellar companion, as previously done for HD\,100546 \citep{2013A&A...557A..68M}. 

\subsection{Caveats and limitations}     
First, we only explored a limited set of dust compositions. Considering crystalline silicates \citep{2010ApJ...721..431J} or other featureless species (e.g., iron) may influence the dust mass and change the disk temperature and brightness profile. However, the available data do not allow us to disentangle more complex dust compositions.\\ 
Our best-fit solution assumed a radially and vertically homogeneous composition. Considering dust radial segregation or settling will affect the disk opacity profile and may lead to different scale heights and surface density profiles that are compatible with the data. However, the available dataset on HD~139614 currently lacks resolved mm observations that probe larger grains and cannot allow us to investigate dust segregation. \\
With a disk orientation close to face-on, the assumption of isotropic scattering may have induced too much NIR flux being scattered in our line of sight and biased our estimation of the inner disk dust mass. However, this should be limited since the NIR scattering opacity is strongly dominated by the sub-$\mu$m-sized grains of our disk model. These grains are still in the Rayleigh regime, in the IR, and scatter almost isotropically.\\   
The 2-D axisymmetric geometry we assumed is relevant for identifying the main disk structural aspects but not possible asymmetries. However, this is not problematic here since the available closure phases do not suggest \changev{strong} brightness asymmetries. We also considered a parameterized dust density that follows the prescription of a gaseous disk in hydrostatic equilibrium. A self-consistent computation of the disk vertical structure, as in \citet{2014A&A...564A..93M}, may have modified our derived dust scale height. However, this approach still relies on a thermal coupling between dust and gas and does not necessarly allow exploring vertical structures departing from the "hydrostatic" prescription. Finally, the power-law surface density we used is an approximation that allows trends in the radial dust distribution to be identified. It remains very simplified compared to the dust density profiles derived from hydrodynamical simulations coupling gas and dust \citep[e.g., ][]{2010A&A...518A..16F} even though the latter can approach power-law profiles in some cases \citep[e.g., ][]{2012A&A...539A.148B}.       
\section{Discussion}
\subsection{A group I Herbig star with a pre-transitional disk}
Our new results on the gapped disk of HD\,139614 support the idea that most of the group-I Herbig objects are in the disk-clearing stage, as suggested by \citep{2013A&A...555A..64M} and confirmed recently by \citet{2015A&A...581A.107M} from a global analysis of all the MIDI data obtained on Herbig stars. Interestingly, \citet{2015A&A...581A.107M} find that Group I objects known to harbor very large gaps of tens up to a hundred  au, such as HD~142527 \citep[e.g., ][]{2006ApJ...636L.153F}, would also have a small gap in the inner region ($\leq 10~$au). Moreover, several supposedly gapless and settled Group~II disks show hints of a very narrow gap in their inner ($\sim $~au) dust distribution. 
\changev{HD\,139614 stands out here} since it is a rare, if not unique, case of Herbig star's disk for which an au-sized gap ($\sim 3$~au) has been clearly spatially resolved in the dust distribution \citep[see][for a review on gap sizes]{2014prpl.conf..497E}. MIR gaps correspond to local depletion in small sub-$\mu$m to $\mu$m-sized grains, which are good tracers of the gas \citep[e.g., ][]{2005A&A...443..185B}. As a result, HD~139614 is a relevant laboratory for investigating gap formation and disk dispersal in the inner regions, which are only reachable by IR interferometry.\\ 
%
%
%
%
%
As shown in Sect.~4, self-shadowing by the inner disk does not convincingly reproduce the pretransitional features of the HD~139614's SED and the interferometric data. Among the main disk clearing mechanisms, two could explain a narrow gapped structure: photoevaporation and planet-disk interaction \citep{2014prpl.conf..497E}. 
The photoevaporation scenario may be relevant for HD~139614 given its accretion rate \citep[$ \lesssim 10^{-8}$~M$_{\odot}$/yr, ][]{2006A&A...459..837G}, which is in the range of predicted photoevaporative mass-loss rates $ 10^{-10}$--$10^{-8}$~M$_{\odot}$/yr \citep{2014prpl.conf..475A}. Moreover, the photoevaporative wind is expected to first open a gap at the critical radius where the mass-loss rate is at its maximum \citep[][]{2014prpl.conf..475A}. In the extreme-UV regime, this critical radius was estimated to be $R_{c,EUV} \simeq 1.8M_*/M_{\odot}$~au. For HD~139614 ($M_*=1.7$~M$_{\sun}$), $R_{c,EUV} \simeq 3.1$~au, which is consistent with the gap location. However, several aspects of the HD~139614 disk hardly appears consistent with this scenario:
\begin{itemize}
 \item the detection of warm molecular gas in the 0.3--15~au region from observations of the rovibrational emission of $^{12}$CO and its isotopologue lines with CRIRES (Carmona et al., in prep.). Preliminary results suggest at least 0.4~M$_{\oplus}$ and up to 2~M$_{\rm Jup}$ of gas within 6.5~au. Moreover, the CO lines present a clear double-peaked profile that is indicative of a gaseous disk in Keplerian rotation, which is not consistent with the presence of a photoevaporative molecular disk wind;
\item the non-detection of the [OI] line at 6300~$\AA$, which suggests the absence of a disk wind \citep[][]{2005A&A...436..209A};
\item the short timescale ($< 10^{5}$~yrs) expected for the inner disk dissipation (in gas and dust) after the photoevaporative wind has opened a gap \citep[][]{2014prpl.conf..475A}. 
\end{itemize}
For these reasons, we then focus on the planet-induced gap scenario. 
Single massive planets are expected to carve a few au wide gap in the gas; the precise width and depth depending on the planet mass and disk viscosity \citep[e.g.,][]{2014prpl.conf..667B}. A similar clearing is expected for the gas-coupled small dust grains. Moreover, the tenuous HD~139614 inner disk suggests a drastic evolution and differentiation of the inner regions, which is possibly reminiscent of dust filtration by a planet on the gap's outer edge \citep[e.g.,][]{2006MNRAS.373.1619R}. Interestingly, \citet{Carmona2014} find similar differences for the HD\,135344\,B transition disk. 
To determine whether our observational constraints on the dust can sustain the planet-induced gap scenario, we performed a comparative study with hydrodynamical simulations of gap opening by a giant planet, adapted to the case of HD~139614.

\subsection{Investigation of disk-planet interaction}

\subsubsection{Hydrodynamical setup} 
\begin{figure*}[t]
 \centering
\resizebox{\hsize}{!}{\includegraphics[width=75mm,height=49mm]{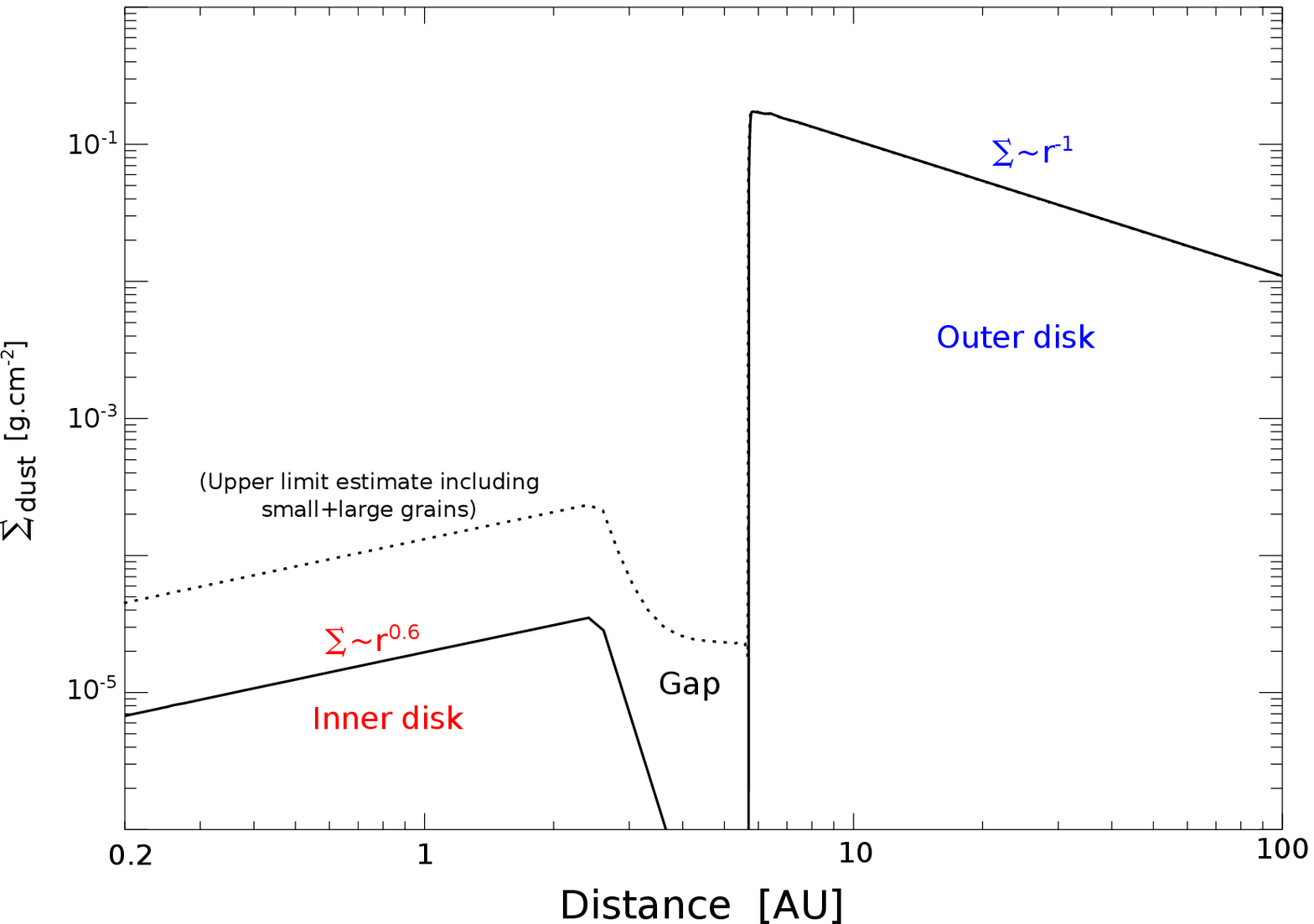}\hspace{0.4cm}
\includegraphics[width=75mm,height=49mm]{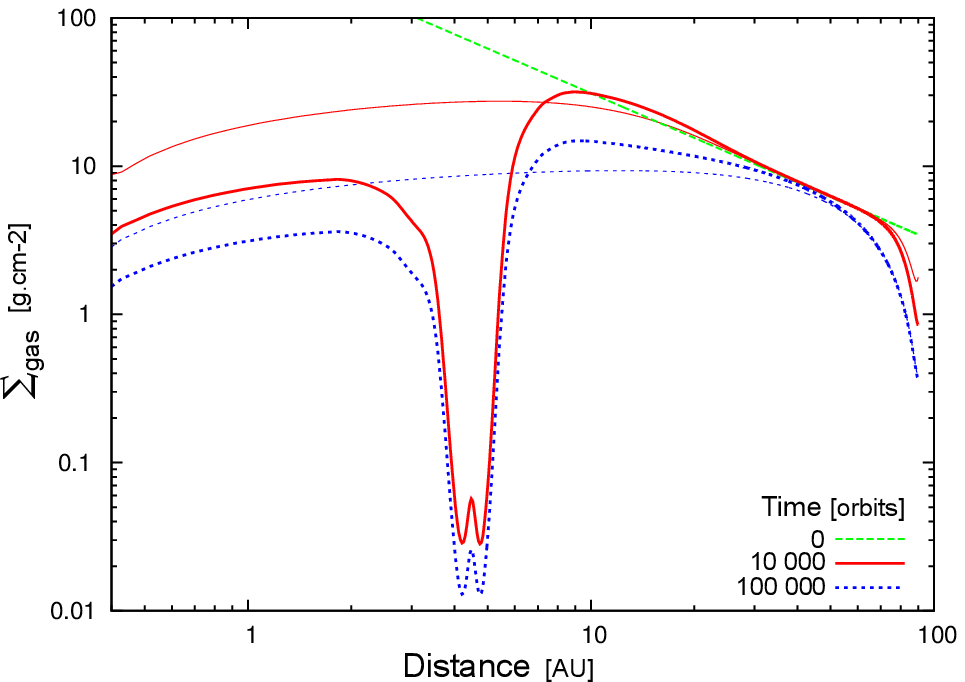}}
  \caption{{\footnotesize Left: Dust surface density profile of the best-fit radiative transfer model as a function of the distance to the star. The upper limit on the dust surface density (small + large grains) in the inner disk and the gap is overplotted (dotted line). Right: Azimuthally averaged gas surface density profiles at various times in a gaseous disk perturbed by a 3.4 M$_{\rm Jup}$ planet. The green line is the initial gas surface density profile; the two thinner and faded solid lines represent the gas surface density profiles after 10\,000 ($\sim 10^5$~Yrs) and 100\,000 orbits ($\sim$1~Myr), without including the planet.}}
 \label{fig:surfacedensity}
\end{figure*}

We use the code FARGO-2D1D\footnote{Available
  at \texttt{http://fargo.in2p3.fr/spip.php?rubrique16}}, which is designed to study the global evolution of a
gaseous disk perturbed by a planet
\citep{2007A&A...461.1173C}.
Our planet is on a circular orbit of radius $a_{\rm pl}$ and does not
migrate. We consider planet masses of 1, 2, and $4\times 10^{-3}\,M_*$, with $M_*$ the mass of the star. For HD~139614, this corresponds to 1.7, 3.4, and 6.8 M$_{\rm Jup}$. The 2D grid extends from $r=0.35\,a_{\rm pl}$ to $2.5\,a_{\rm pl}$ with
$N_r=215$ rings of $N_s=628$ cells arithmetically spaced so that the resolution is ${\rm d}\phi={\rm d}r/r=0.01$ at the planet location, where $\phi$ is the azimuth. The
1D grid extends from $0.04\,a_{\rm pl}$ to $20\,a_{\rm pl}$ and has open boundary
conditions to allow for disk spreading.
The simulation uses a locally isothermal equation of state:
$P={c_S}^2\Sigma$ with $P$ the pressure, $\Sigma$ the gas surface
density, and $c_s=H\Omega$ the sound speed. Here, $c_s$ is fixed
such that $H/r$ is constant, with $H$ the gas pressure scale height and $\Omega$  the Keplerian angular velocity. We set $H/r=0.04$ to be consistent with the $H/r$ value inferred for the dust at the outer disk's inner edge. The disk scale height influences the gap's depth but is not a critical parameter here, given our limited constraint on the dust gap depth. The initial density profile is
$\Sigma=\Sigma_0(r/a_{\rm pl})^{-1}$. The gas viscosity is
$\nu=10^{-5}\sqrt{GM_*a_{\rm pl}}$, with $G$ the gravitational constant; at the planet location, the Reynolds number is
$R=r^2\Omega/\nu=10^5$, which gives $\alpha=6.25\times
10^{-3}$ in the \citet{1973A&A....24..337S} prescription, but
here the viscosity is assumed to be independent of time and space.
With this equation of state and the planet on a fixed orbit,
the units are arbitrary: Regardless of the physical value of $a_{\rm pl}$ and
$\Sigma_0$, the results scale accordingly. For easier comparison with the case of HD\,139614, we have set $a_{\rm pl}=4.5$~au, and $\Sigma=15$~g cm$^{-2}$ at 10~au as extrapolated from the dust surface density at 10~au ($\Sigma_{\rm dust}=0.15$~g cm$^{-2}$), assuming a gas-to-dust ratio of 100. Figure \ref{fig:surfacedensity} shows the gas density profiles.
\subsubsection{Simulated gas density profile}
With the considered disk viscosity and scale height, a $\sim 3$ M$_{\rm Jup}$ planet at 4.5~au produces a gas gap between $\sim$ 3 and 6~au. This is consistent with the dust gap width, the inner disk's outer radius ($r_{\rm out}=2.55\pm0.13$~au), and the outer disk's inner radius ($r_{\rm in}=5.7$~au) of our best-fit dust model. The gas gap appears shallower than in the dust, which is expected since gas pressure minima are strongly depleted in dust \citep[e.g., ][]{2006MNRAS.373.1619R}. 
The 1.7 and 6.8~M$_{\rm Jup}$ planets produced a gap that was either too narrow ($\lesssim 2$~au) or too wide ($\geq 4$~au).\\
\noindent Another noticeable aspect is the surface density profile. From the radiative transfer, we highlighted a radially increasing dust surface density profile in the inner disk, while the usual profile in $r^{-1}$ was kept for the outer disk. The gas surface density profile appears very consistent with this. In the inner disk, the gas surface density increases from $0$ at the inner edge until it reconnects to the original profile in $r^{-1}$ at $r>5$~au. Moreover, if fitted by a power law, the gas profile's exponent in the inner disk ($\sim 0.6$) is similar to the dust profile exponent ($0.64$). This radially increasing gas surface density, which we highlighted for the small dust grains, too, is intrinsic to the disk viscous evolution. As shown in \citet{1974MNRAS.168..603L} and \citet{2007MNRAS.377.1324C}, such a profile naturally appears in the inner region of a viscously spreading gaseous disk, which has a finite inner edge that is not too close to the star (possibly $4\%$ of the  planet's orbital radius for HD~139614).
Interestingly, a radially increasing gas density profile was required to describe the CO emission within the cavity of the HD\,135344B disk \citep{Carmona2014}.\\
After a gap has been opened, the gas surface density ratio between the inner and outer components is mainly ruled by the amount of gas that can cross the gap. 
Moreover, if the gap is opened close to the disk's innermost edge, where the density profile is radially increasing, the inner disk's surface density will be lower than in the outer disk \citep[see Fig.6 of ][]{2007MNRAS.377.1324C}.  
Figure \ref{fig:surfacedensity} shows an inner disk surface density deficit of a factor 10 in gas, relative to the outer disk. This does not vary much as the whole disk viscously spreads and the gas density decreases as a whole. Even after 1~Myr, the inner disk is much less depleted in gas ($\sim 10$) than in dust ($\sim 10^3$). This suggests that a planet opening a gap consistent with our observations cannot affect the accretion flow enough to the inner regions. With the 6.8~M$_{\rm Jup}$ planet or more viscous disks, we could not deplete the inner disk any more. Having the planet on a fixed
orbit actually helps the inner disk's depletion since a planet
migrating in type-II migration follows the disk viscous spreading without hampering it. Also, when the disk is less massive than the
planet, no migration occurs, and the situation is similar to our simulation \citep[see Eq.11 of ][]{2007MNRAS.377.1324C}. 
However, during the disk evolution, dust grains grow and decouple from the gas, or fragment and stay coupled. The inner disk depletion may thus differ for the small dust grains and the gas, as shown hereafter.
\subsubsection{Grain growth and fragmentation in the inner disk}
The outer edge of a gap opened by a planet is a pressure maximum and can trap the partially gas-decoupled dust grains. In the inner disk, these grains drift inward and are eventually lost to the star. The radial drift velocity is highest when the Stokes number $\mathrm{St}=\Omega\rho_\mathrm{d}a/\rho_\mathrm{g}c_\mathrm{s} \sim 1$,
where $\Omega$ is the Keplerian angular velocity, $\rho_\mathrm{d}$ the intrinsic dust density, $a$ the grain size, $\rho_\mathrm{g}$ the gas density, and $c_\mathrm{s}$ its sound speed. The optimal size for radial migration is $a_\mathrm{opt}=\Sigma_\mathrm{g}/\sqrt{2\pi}\rho_\mathrm{d}$
in the midplane \citep{2010A&A...518A..16F}. Since the gap acts as a dust filter, the dust-to-gas ratio in the outer disk is likely to be close to the interstellar value of 0.01, leading to $\Sigma_{\rm g}\sim10$~g\,cm$^{-3}$ at the gap's outer edge (see Fig.~\ref{fig:surfacedensity}). 

\noindent Our simulations indicate a gas surface density that is ten times lower in the inner disk, which implies $a_\mathrm{opt} \simeq 1$~mm for typical values of $\rho_{\rm d}$ (1--3~g\,cm$^{-3}$). 
Grain growth is expected to be efficient in the inner disk regions \citep{2008A&A...480..859B,2008A&A...487..265L} and to bring small grains to mm or cm sizes. Once grains have grown up to sizes $\sim a_\mathrm{opt}$ in the inner disk, they migrate inwards quickly and sublimate when approaching the star. This could occur on a timescale of $10^2$~yr for cm-sized bodies at 1~au in a typical accreting disk \citep{2008A&A...480..859B}. Only the small grain ($\mu$m-sized or smaller) reservoir, partly replenished by fragmentation, would stay coupled to the gas and remain in the inner disk longer. This depletion of the small grains reservoir will lead to a tenuous inner disk with reduced optical depth.
This is supported by hydrodynamical simulations of a disk of gas and dust containing a planet, where dust grain growth, fragmentation, and dynamics are treated self-consistently \citep{Gonzalez2015}. Although the disk and planet parameters are different from ours, they show an inner disk that depletes more efficiently in dust than in gas. For most of the considered
fragmentation velocities ($<20$~m\,s$^{-1}$), growth is fast for small grains but more difficult as they keep growing. Fragmentation either prevents grains from growing above $a_\mathrm{opt}$, promoting their fast migration, or breaks them down again to smaller sizes, thereby preventing their fast migration and partly replenishing the small grain reservoir. The inner dust disk thus drains out on timescales shorter than the viscous disk evolution and only keeps the remaining small grains and the larger bodies that could have overcome the radial-drift and fragmentation barriers. After $10^5$~yr, the inner-to-outer disk dust density ratio has dropped to $\sim10^{-3}$ and leads to a tenuous inner disk with a reduced optical depth, which agrees with our observational results.
Quantitatively reproducing our observed dust disk with similar simulations is beyond the scope of the paper. Nevertheless, the qualitative agreement supports the scenario of a planet-induced gap in the HD\,139614 dust disk.
\section{Conclusion}
This first multiwavelength modeling of the dust disk around HD~139614 provided the following results: 
\begin{itemize}
\item We confirmed a gap structure, between 2.5 and 5.7~au, depleted \changev{(but not necessarily empty)} in warm sub-$\mu$m- and $\mu$m-sized grains. HD~139614 appears to be a rare case of a Herbig star with a narrow au-sized gap in its dust disk. 
\item The NIR-emitting region was found to be spatially extended from 0.2 to 2.5~au, with a radially increasing surface density profile ($p>0$), a dust scale height of $H\sim0.01$~au at $r=0.2$~au, and a strong dust depletion of at least $\sim 10^3$ (in surface density) relative to the outer disk. This suggests a drastic evolution of the inner regions. 
\item  With a dispersion of $\lesssim 2-5^{\circ}$ around zero, the PIONIER closure phases does not suggest significant brigthness asymmetries. Notably, HD\,139614 is unlikely to be a tight binary system with a companion more massive than about 0.11~M$_{\odot}$.
\end{itemize}
Currently available data and modeling do not support self-shadowing from the inner disk or photoevaporation as the origin of the dust gap in the HD~139614 disk. We thus tested whether the mechanism of disk/planet interaction could be viable against our constraints obtained on the dust from the radiative transfer. 
Assuming that small dust grains, probed by IR interferometry, are coupled to the gas, we performed hydrodynamical simulations of gap formation by a planet in a gaseous disk using the code FARGO-2D1D. It appears that:
\begin{itemize}
 \item For typical values of disk viscosity and scale height, a 3~$M_{\rm jup}$ planet, \changev{fixed} at $\sim$4.5 au, could produce a gap in the gas consistent with the dust gap, although a bit shallower. 
 \item The radially increasing dust surface density in the regions interior to the gap is reproduced in the gas. The original gas density profile ($\propto r^{-1}$) is maintained in the outer disk for a long time. The radially increasing profile is predicted analytically and represents the accretion profile of the gas (and coupled dust grains) close to the inner disk rim. To our knowledge, it is the first time that this behavior is observationally identified in the innermost dust distribution of a disk.  \item The inner disk dust depletion ($\sim 10^{-3}$) is not reproduced in gas ($\sim 0.1$). However, this can be explained by the radial drift and growth and fragmentation processes affecting dust and is predicted by hydro simulations coupling gas and dust.
 \end{itemize} 
 All of this supports the hypothesis of a planet-induced gap in the HD\,139614's dust disk
and reinforces the idea that disks around Group-I Herbig stars are already in the disk-clearing transient stage. Confirming the planet-induced gap scenario will require further observations such as upcoming MIR imaging of the gap with the second-generation VLTI instrument MATISSE \citep{2014Msngr.157....5L}. Moreover, ALMA will be able to probe the mm grains and obtain a robust estimate of their mass at $r < 10$~au, while mapping the gas density distribution and dynamics inside and outside the gap. Although the HD\,139614 gap seems out of reach for VLT/SPHERE, direct imaging of the outer regions could be attempted to detect other signatures of disk-planet interaction. HD\,139614 will also constitute an exciting target for the future E-ELT/MICADO, PCS, and METIS instruments.

\section*{Acknowledgments}
We thank the anonymous referee for the comments that helped to improve this manuscript significantly.
  A. Matter acknowledges financial support from the Centre National
  d'{\'E}tudes Spatiales {\small (CNES)}. J.-F.~Gonzalez is grateful to the LABEX Lyon Institute of Origins
{\small (ANR-10-LABX-0066)} of the Universit\'e de Lyon for its financial support
within the program ``Investissements d'Avenir'' {\small (ANR-11-IDEX-0007)} of the
French government operated by the ANR. This work is supported by the French ANR POLCA project {\small (Processing of pOLychromatic interferometriC data for Astrophysics, ANR-10-BLAN-0511)}. This work was supported by the Momentum grant of the MTA CSFK Lend\"ulet Disk Research Group.

\bibliographystyle{aa}

\bibliography{biblioHD139614}

\Online

\begin{appendix}
\section{Parameter effects in radiative transfer modeling}
In this section, we illustrate the individual effect of the main disk parameters on our radiative transfer modeling. From our best-fit model shown in Table~\ref{tab:global} and Figs.~\ref{fig:radmcsed} and \ref{fig:radmcmodel}, we vary each parameter separately around its best-fit value, and show the impact on the modeled observables (SED, NIR $V^2$, and MIR visibilities). In this way, we illustrate the effects of the main disk parameters and how much they can be constrained by our available dataset.
\subsection{Inner disk's inner radius}
\begin{figure*}[h]
 \centering
 \resizebox{\hsize}{!}{\includegraphics[width=60mm,height=40mm]{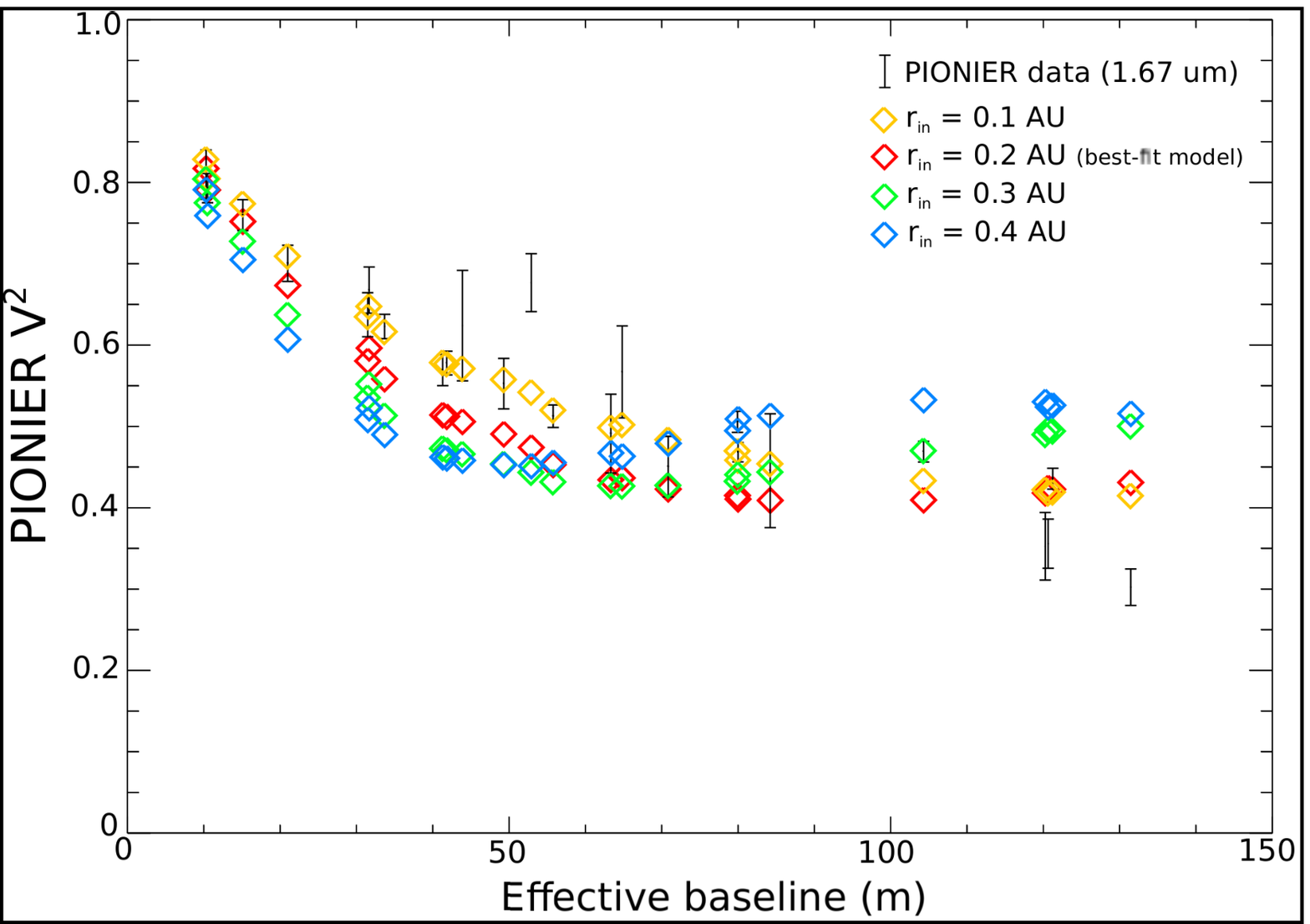}
 \includegraphics[width=60mm,height=40mm]{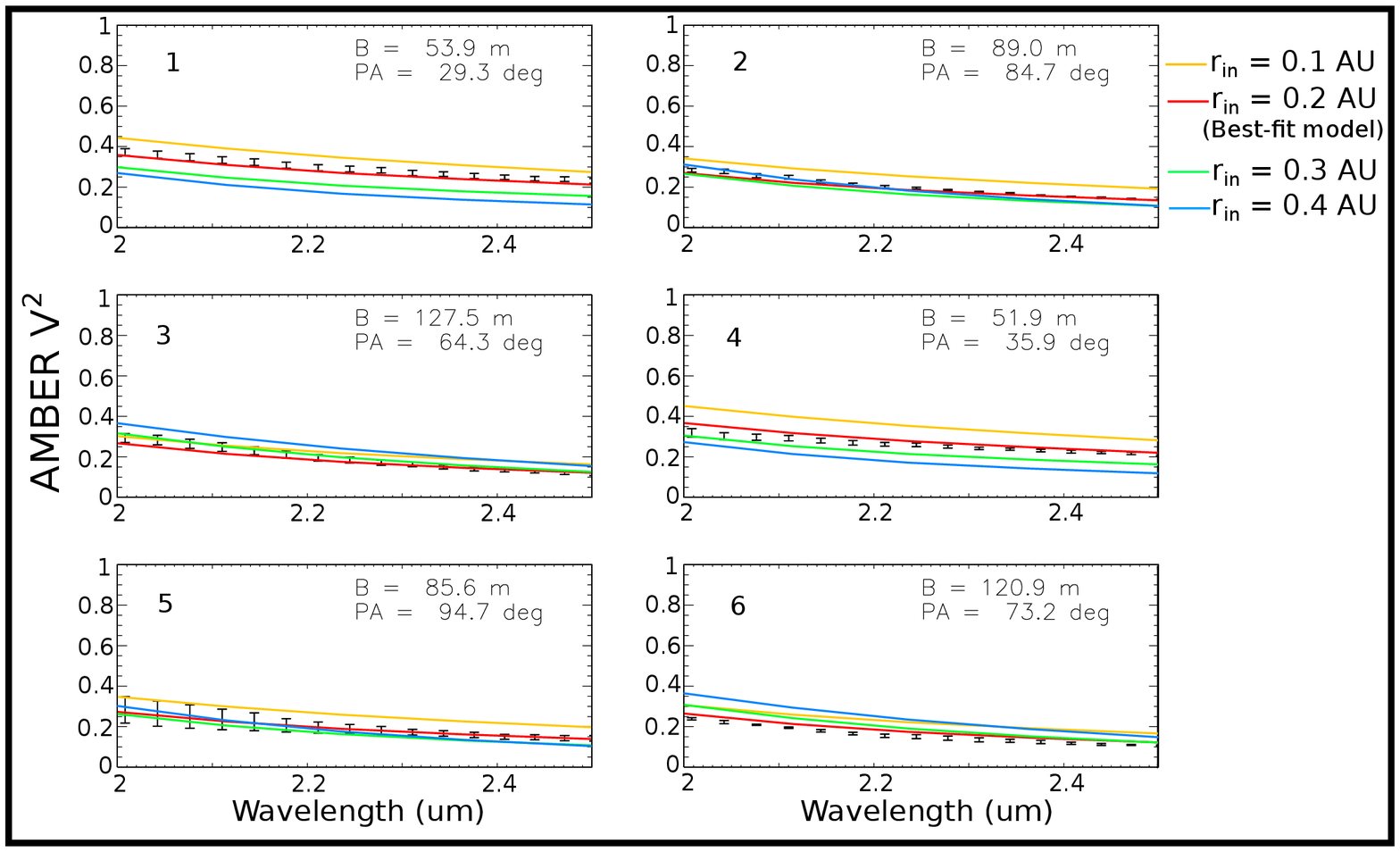}
 \includegraphics[width=60mm,height=40mm]{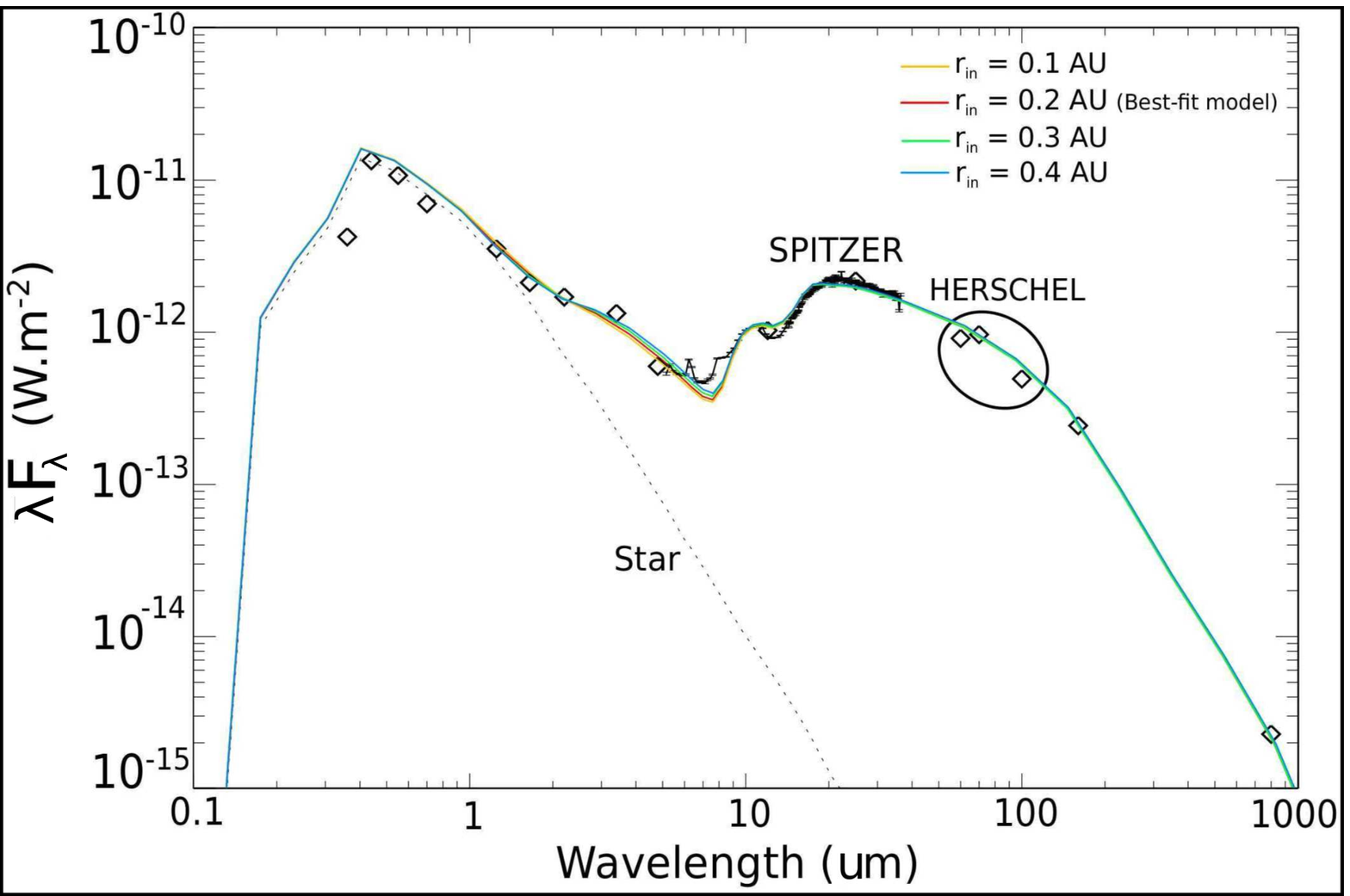}}
  \caption{{\footnotesize Effect on a change of the inner disk's inner radius $r_{\rm in}$. We show the modeled {\itshape H}-band $V^{2}$ at 1.67~$\mu$m (left), {\itshape K}-band $V^2$ (middle), and SED (right), overplotted on the measured PIONIER $V^2$, AMBER $V^2$, and SED, respectively. The best-fit model is shown in red.}}
 \label{fig:rin}
\end{figure*}
We show here the effect of varying the inner radius of the inner disk on the SED and the {\itshape H}-band and {\itshape K}-band $V^2$. As mentioned in Section 4.3.1, $r_{\rm in} < 0.2$~au values slightly
increase the amount of unresolved inner disk emission. As shown in Fig.~\ref{fig:rin}, this induces higher modeled {\itshape H}-band $V^2$ combined with a more gradual $V^2$ decrease as a function of spatial frequency. However, the modeled {\itshape K}-band $V^2$ become overestimated at the lowest spatial frequencies (smallest baselines). The $r_{\rm in} > 0.2$~au values induce too prominent a lobe in the PIONIER $V^2$ profile at high spatial frequencies. The modulation amplitude thus increases too much as we increase $r_{\rm in}$ above 0.2~au. 

\subsection{Surface density $p$ exponent}
\begin{figure*}[h]
 \centering
 \resizebox{\hsize}{!}{\includegraphics[width=60mm,height=40mm]{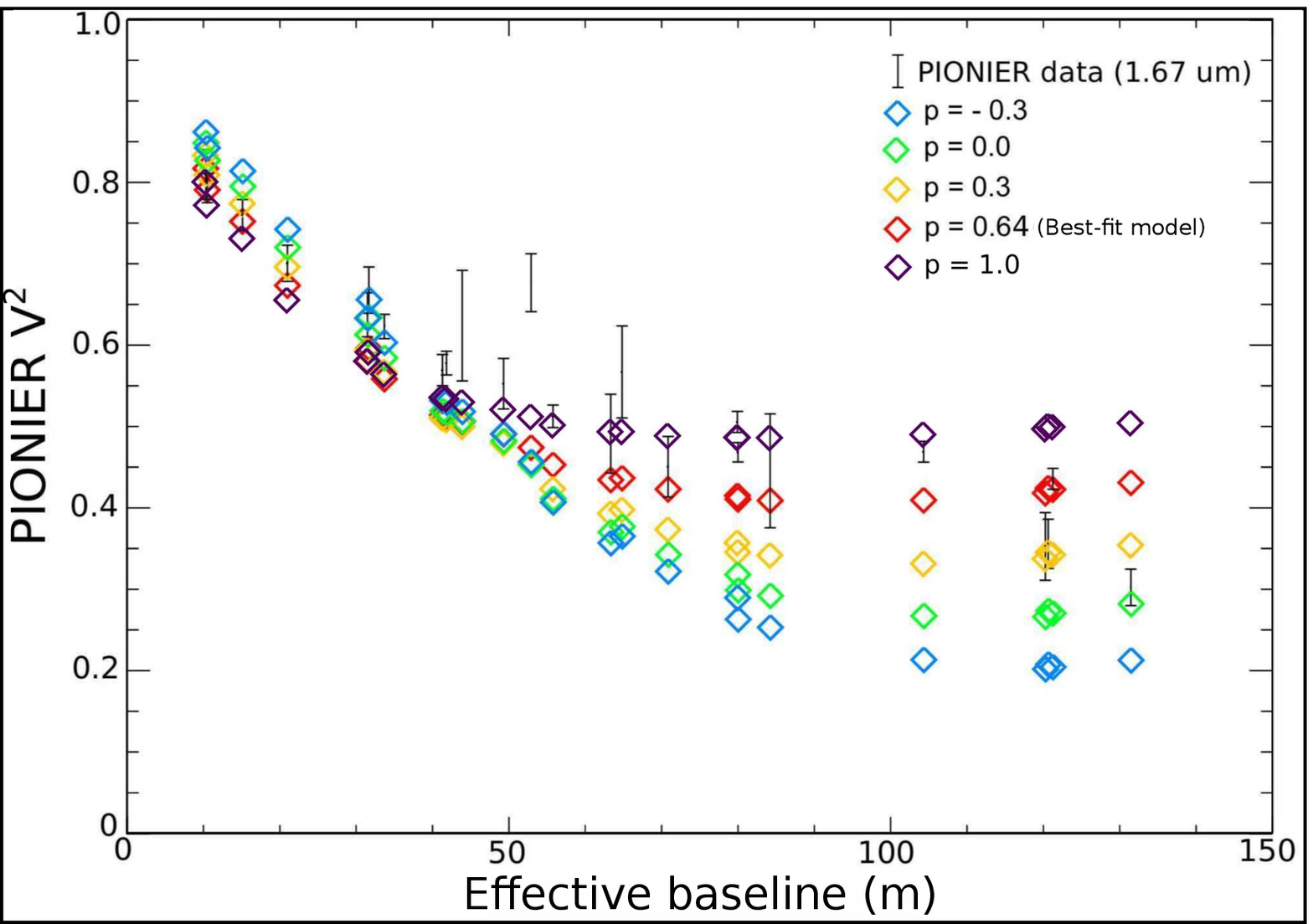}
 \includegraphics[width=60mm,height=40mm]{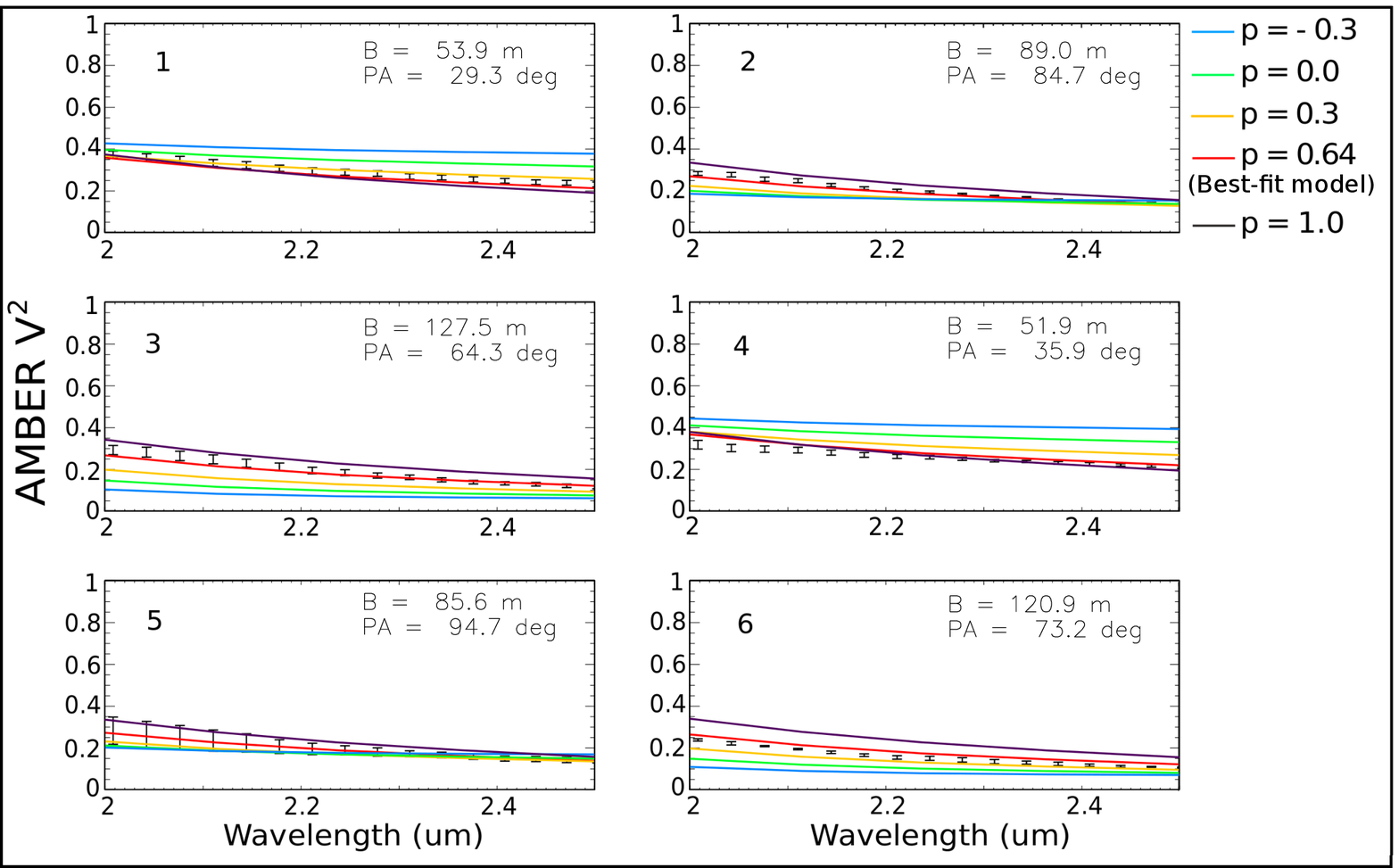}
 \includegraphics[width=60mm,height=40mm]{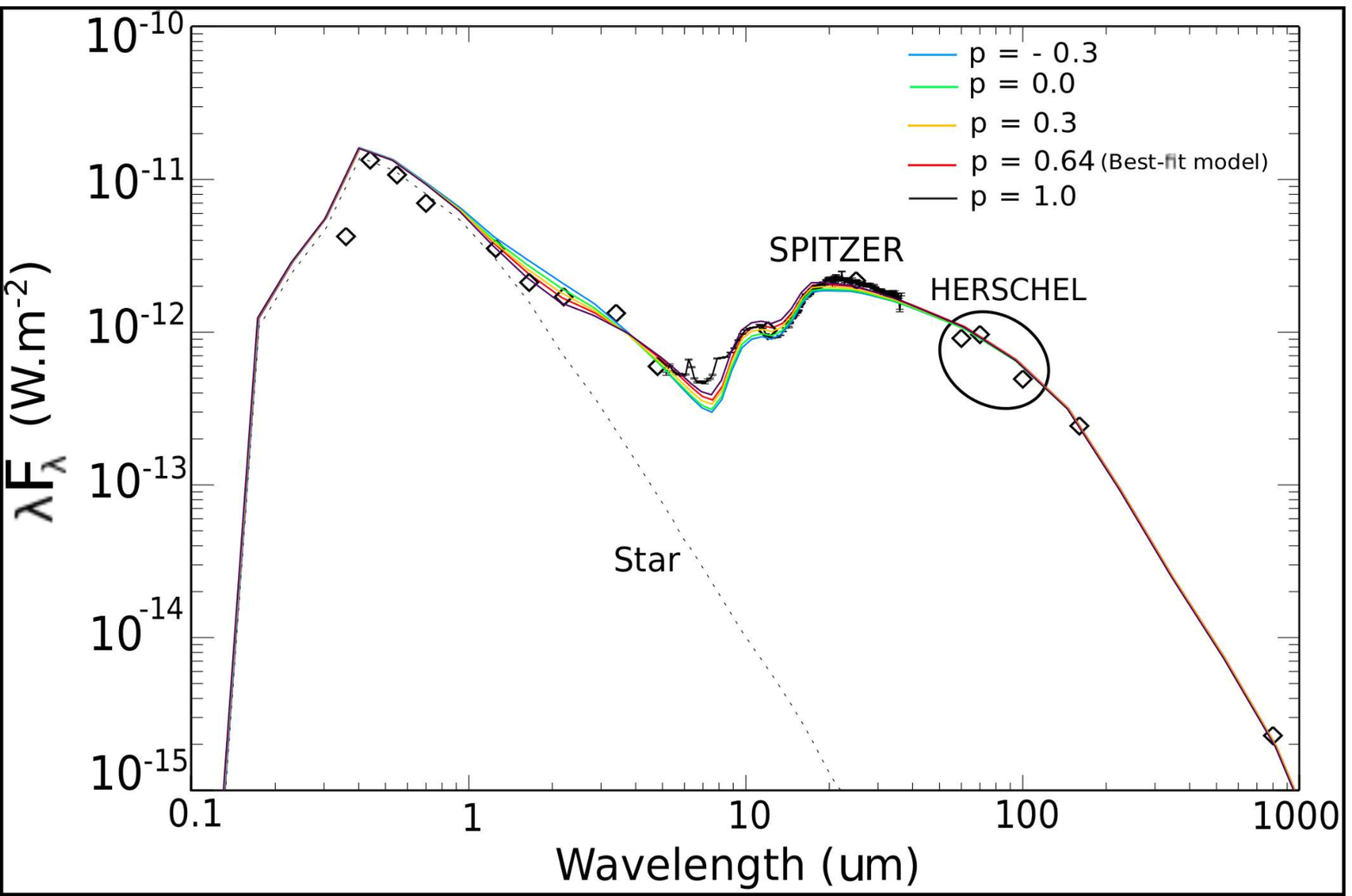}}
  \caption{{\footnotesize Same as in Fig.~\ref{fig:rin} but focusing on the effect on a change of the $p$-exponent of the power-law profile of the inner disk's dust surface density.}}
 \label{fig:p}
\end{figure*}
We examine here the effect of varying the power-law exponent $p$ of the inner disk's dust surface density on the NIR SED and $V^2$.
As mentioned in Section 4.3.1 and shown in Fig.~\ref{fig:p}, the models with $p \leq 0.3$ start inducing too strong a flux contribution from the inner disk rim and an overestimated NIR emission at $\lambda \lesssim 2$~$\mu$ in the modeled SED. This too spatially confined NIR-emitting region implies {\itshape H}-band and {\itshape K}-band $V^2$ that is
too high at low spatial frequencies and $V^2$ that is  too low at high spatial frequencies since the stellar-to-total flux ratio is lower than in the case of our best-fit model, especially at $\lambda \lesssim 2$~$\mu$.
\subsection{Inner disk's outer radius}

\begin{figure*}[t]
 \centering
 \resizebox{\hsize}{!}{\includegraphics[width=90mm,height=60mm]{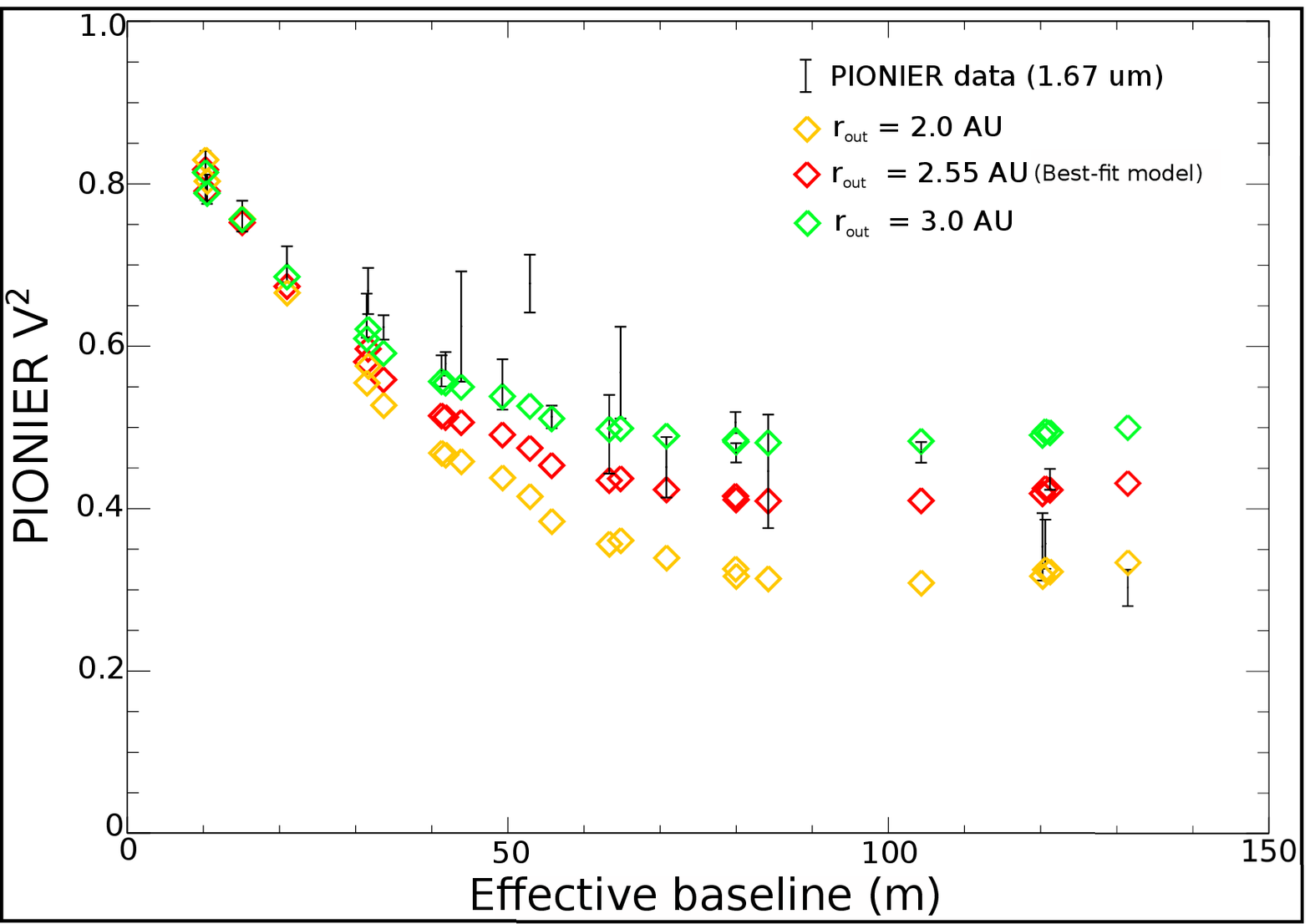}
 \includegraphics[width=90mm,height=60mm]{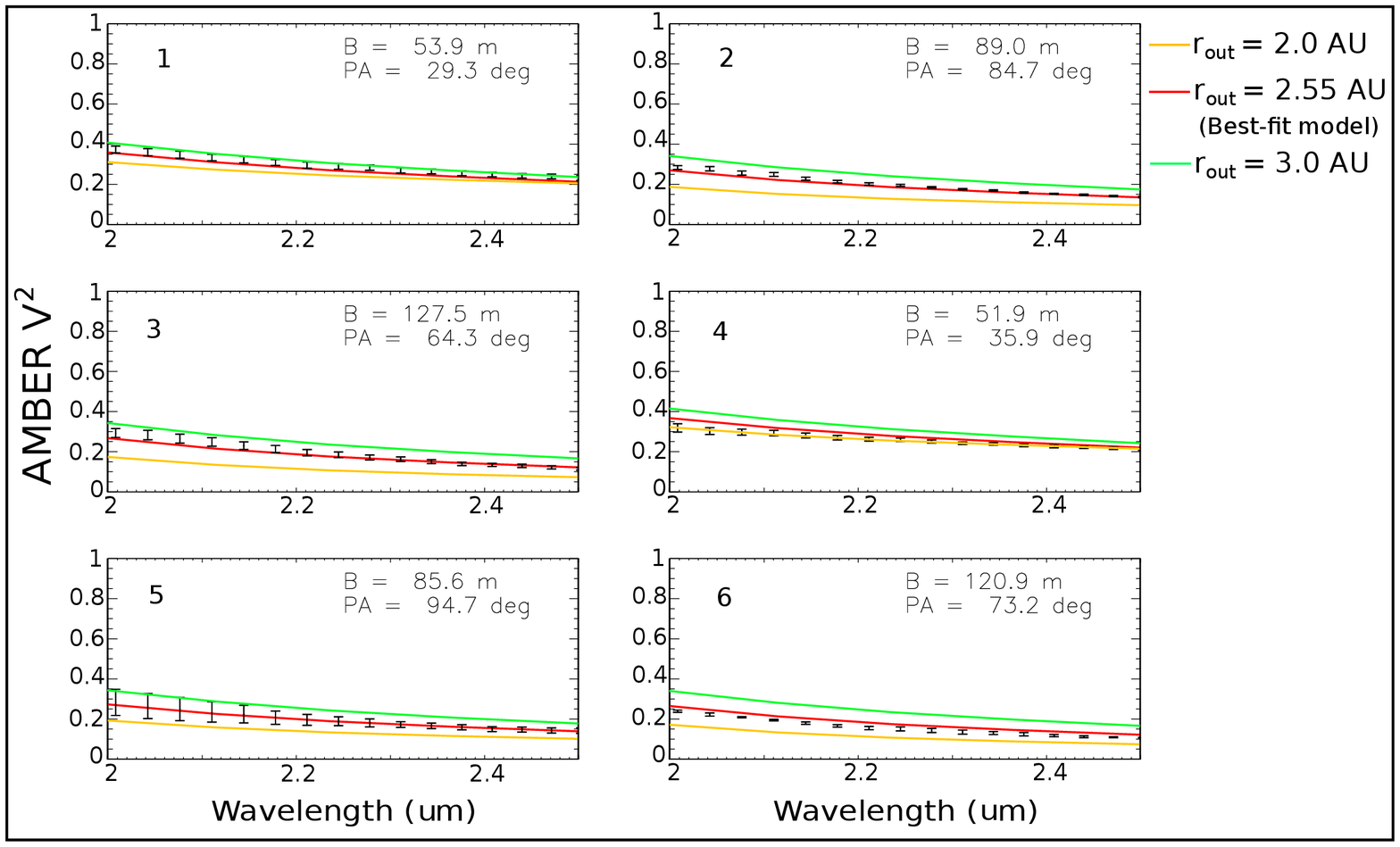}}\\[1.5mm] 
 \resizebox{\hsize}{!}{\includegraphics[width=90mm,height=60mm]{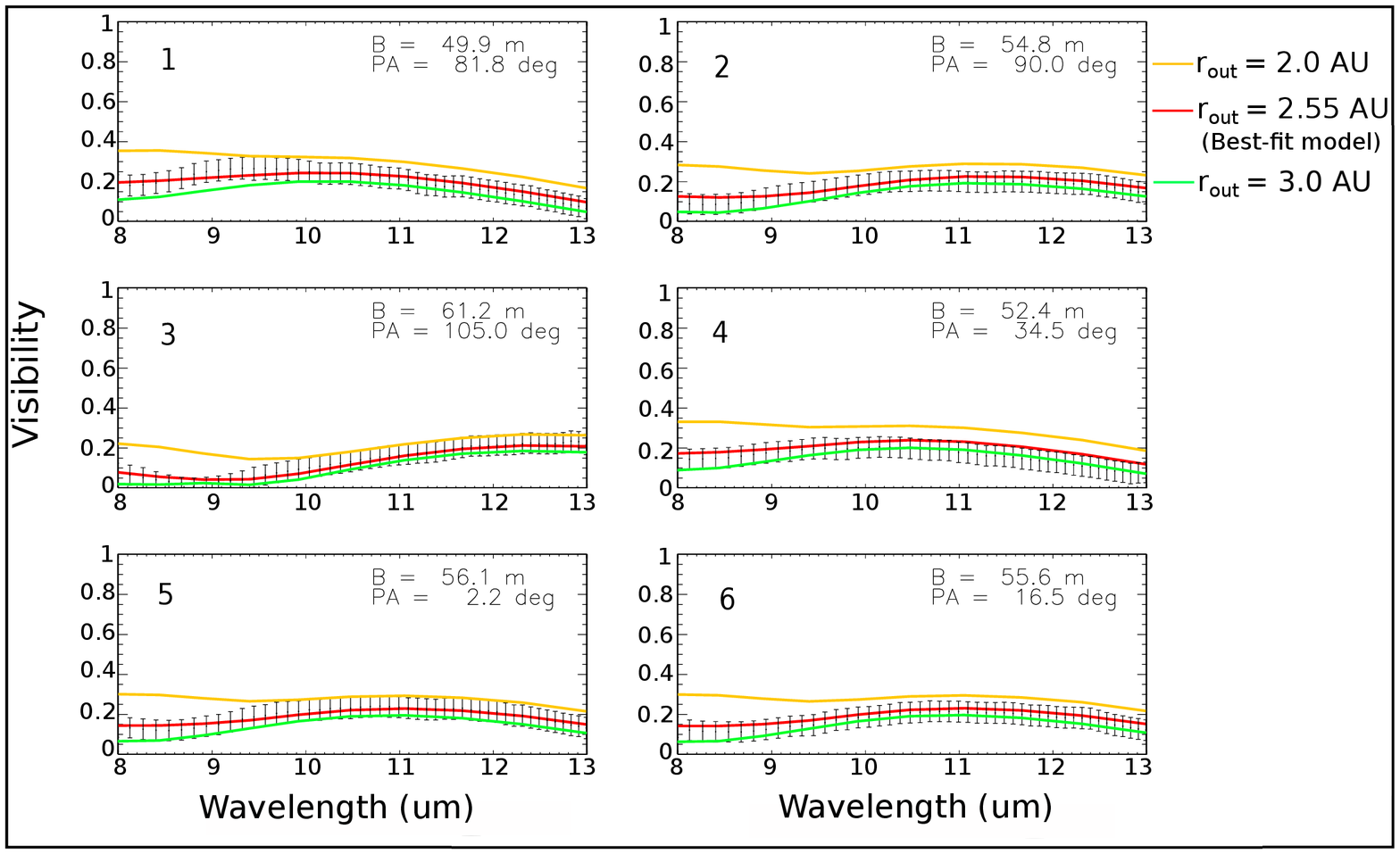}
 \includegraphics[width=90mm,height=60mm]{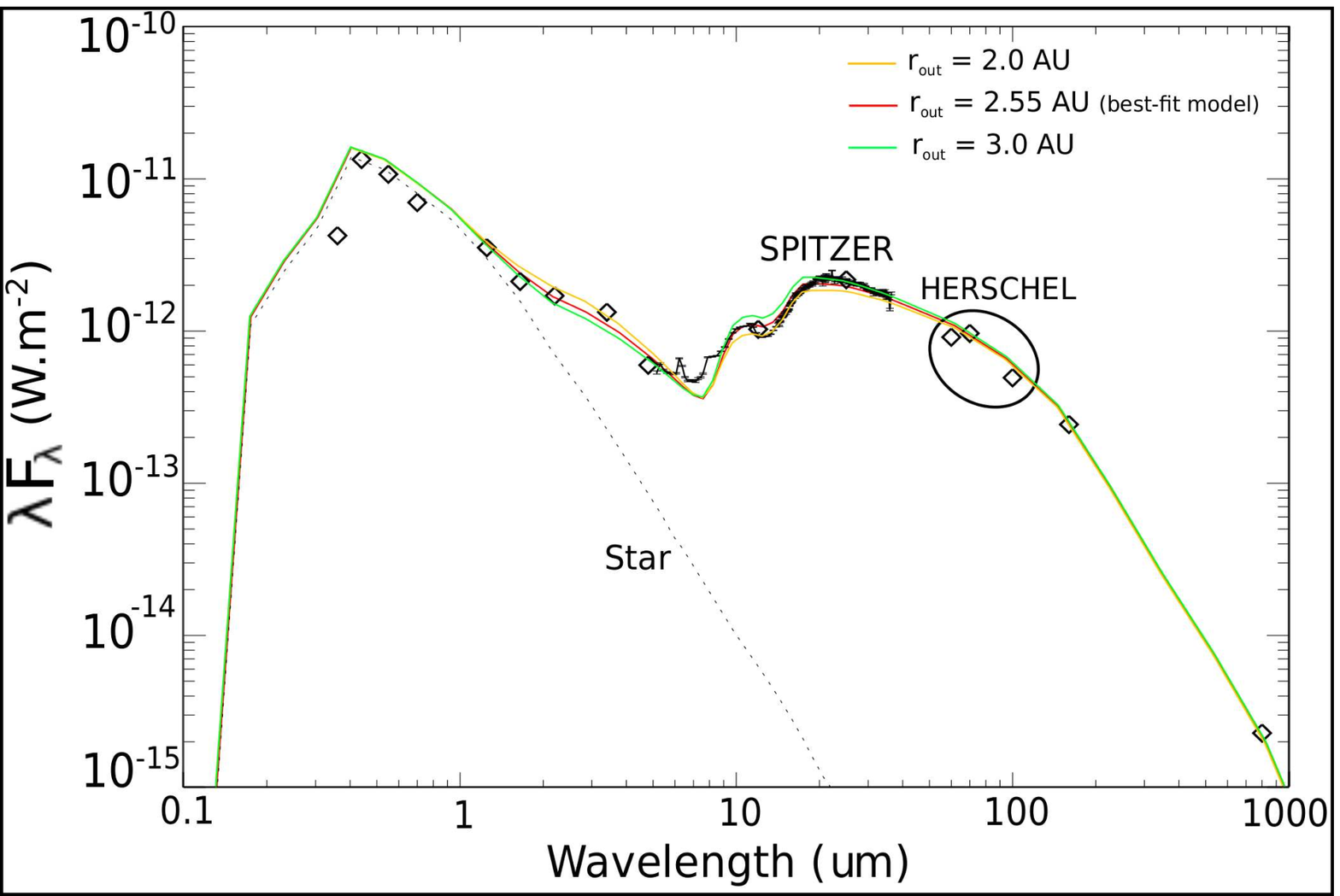} }
  \caption{{\footnotesize Same as in Fig.~\ref{fig:rin} but focusing on the effect on a change in the outer radius of the inner disk $r_{\rm out}$. We also included the modeled {\itshape N}-band visibilities, overplotted on the measured MIDI visibilities (bottom left).}}
 \label{fig:rout}
\end{figure*}

We examine here the effect of varying the outer radius $r_{\rm out}$ of the inner disk. In addition to the NIR SED and $V^2$, we also consider that the {\itshape N}-band visibilities illustrate the constraint provided by the MIDI data on the extension of the inner disk. Figure~\ref{fig:rout} shows that decreasing or increasing the size of the inner disk by at least 0.5~au, relative to our best-fit model, has a noticeable effect on all the observables. A smaller inner disk will induce too high {\itshape N}-band visibilities especially between 8 and 9.5~$\mu$m, while larger and over-resolved inner disks ($r_{\rm out} \geq 3$~au) will produce {\itshape N}-band visibilities that are too low and flatter. The effect on the {\itshape H}-band and {\itshape K}-band $V^2$ is directly correlated to the change in the NIR emission level in the modeled SED. Indeed, a smaller inner disk ($r_{\rm out} \lesssim 2.0$~au) is more optically thick (for the same dust mass) and will produce more NIR emission. This implies an overestimation of the disk contribution and consequently a lower stellar-to-total flux ratio, which will then induce lower NIR $V^2$ (see Fig.~\ref{fig:rout}), especially at long baselines where the inner disk is fully resolved. For a larger inner disk ($r_{\rm out} \geq 3$~au), the effect on the observables is the opposite.

\subsection{Inner disk scale height}
\begin{figure*}[h]
 \centering
 \resizebox{\hsize}{!}{\includegraphics[width=60mm,height=40mm]{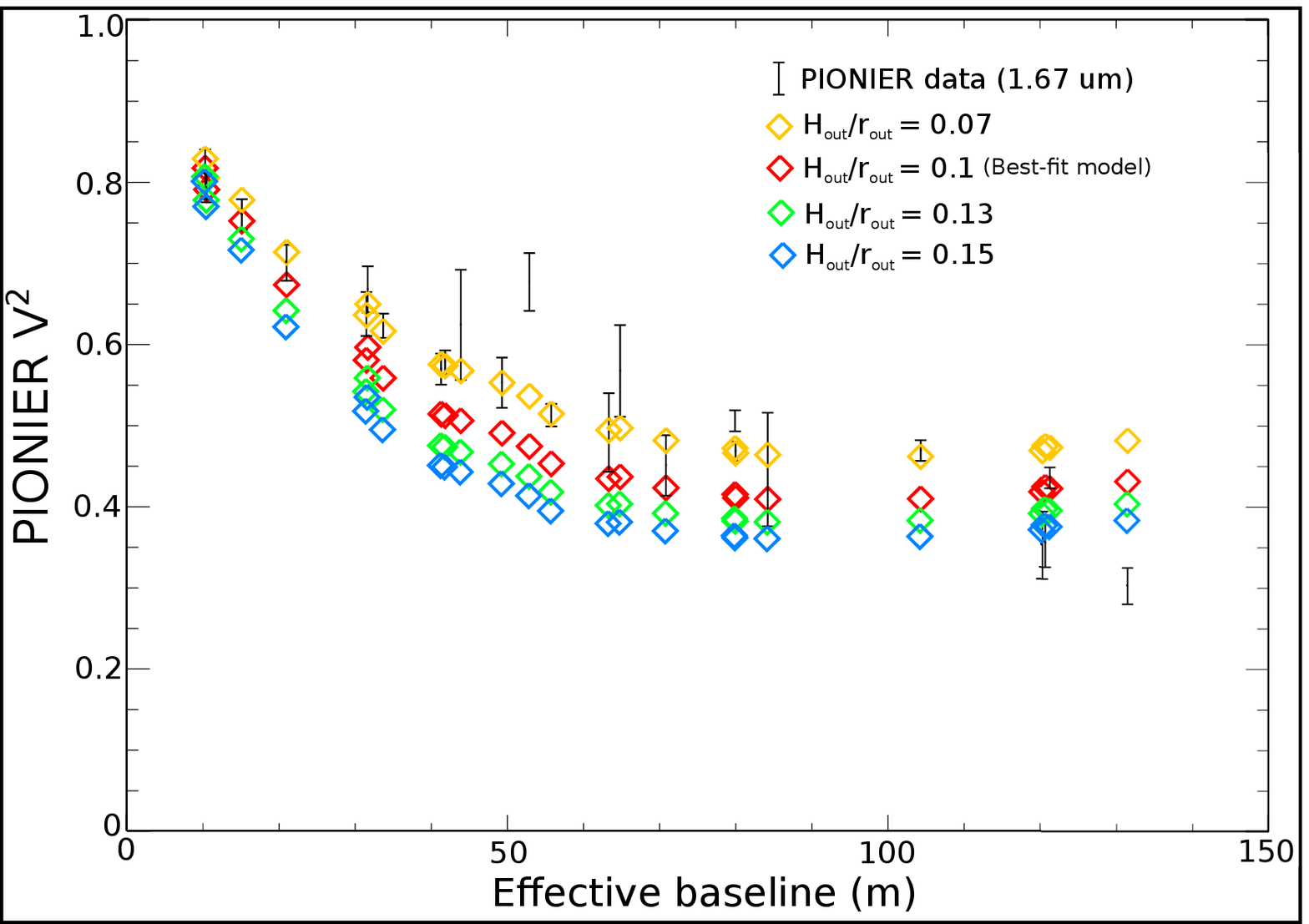}
 \includegraphics[width=60mm,height=40mm]{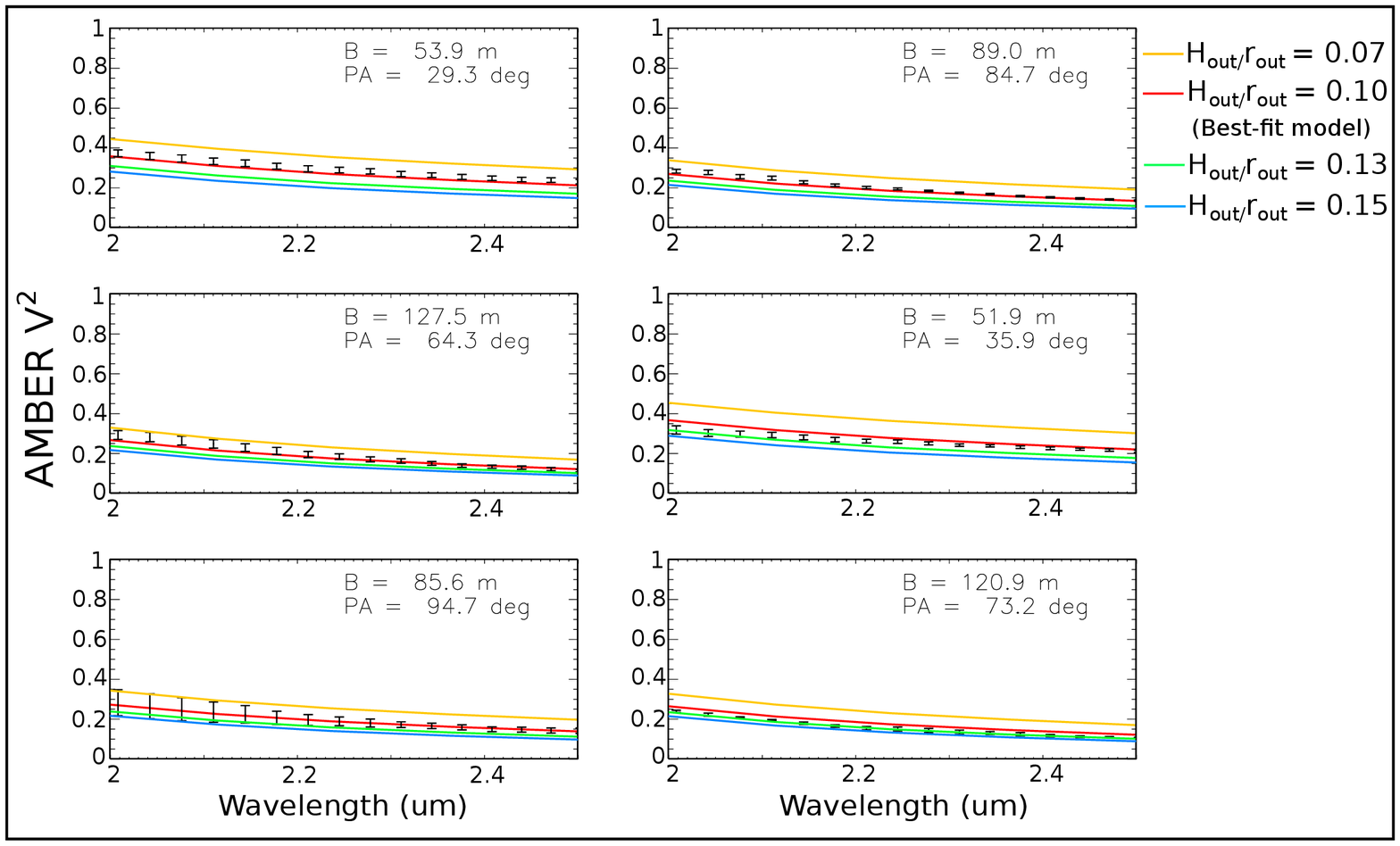}
 \includegraphics[width=60mm,height=40mm]{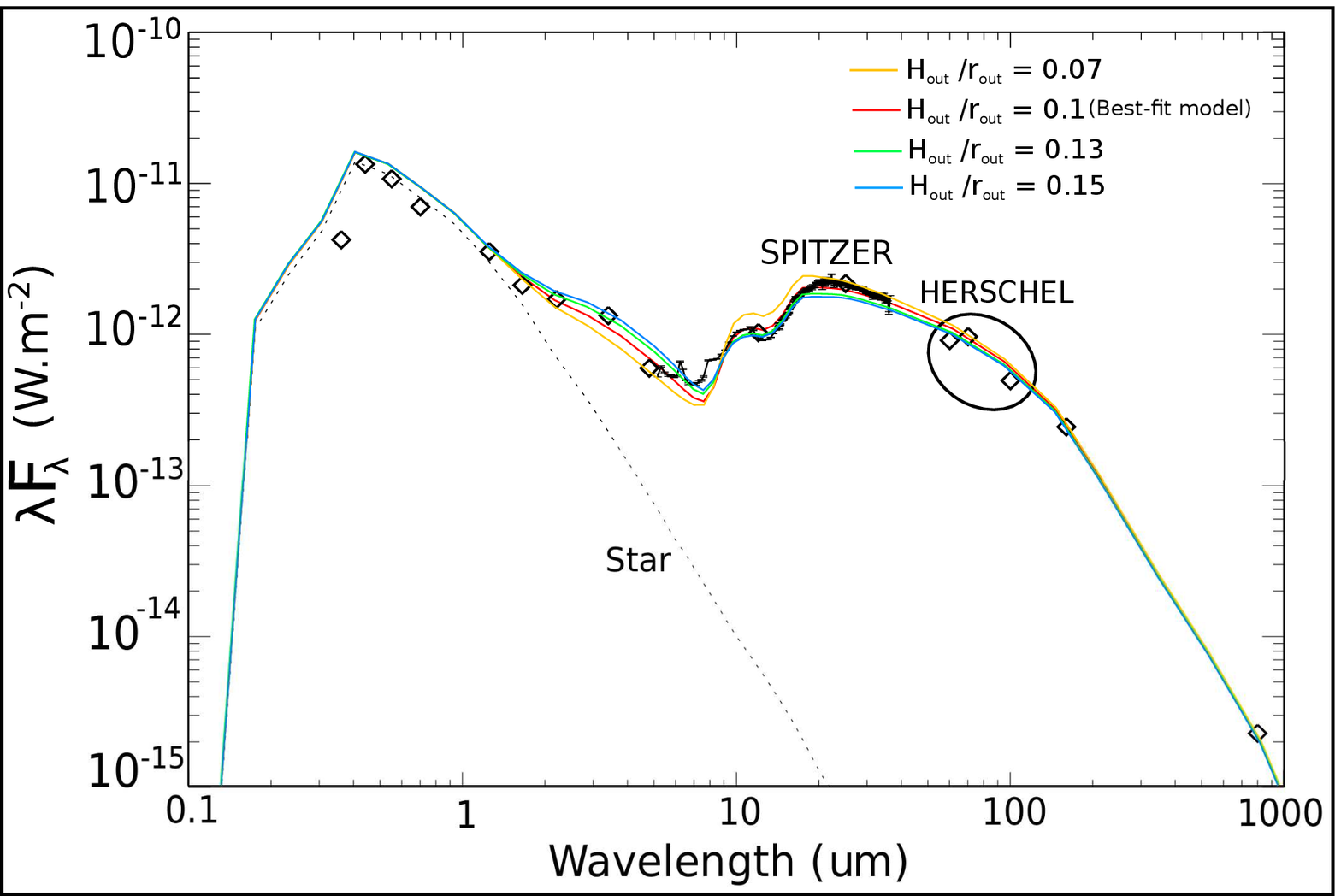}}
  \caption{{\footnotesize Same as in Fig.~\ref{fig:rin} but focusing on the effect on a change in the dust scale height at the inner disk's outer radius $H_{\rm out}/r_{\rm out}$.}}
 \label{fig:h}
\end{figure*}

We show in Fig.~\ref{fig:h} the effect of varying the dust scale height $H_{\rm out}$ of the inner disk at its outer radius $r_{\rm out}$. A slight departure from the best-fit value $H_{\rm out}/r_{\rm out}=0.1$ has a noticeable impact on all the NIR observables. Indeed, increasing the scale height increases the amount of stellar flux captured and reprocessed by the inner disk and thus the NIR emission level. This implies a decrease in the stellar-to-total flux ratio and therefore lower NIR $V^2$, especially at high spatial frequencies. We thus quickly underestimate the $V^2$ level in {\itshape H}  and {\itshape K} bands. On the other hand, slightly reducing the dust scale height to $H_{\rm out}/r_{\rm out} \simeq 0.07$ produces modeled SED and NIR $V^2$ that are still consistent with the observations.

\subsection{Outer disk's inner radius}
\begin{figure*}[h]
 \centering
 \resizebox{\hsize}{!}{\includegraphics[width=90mm,height=60mm]{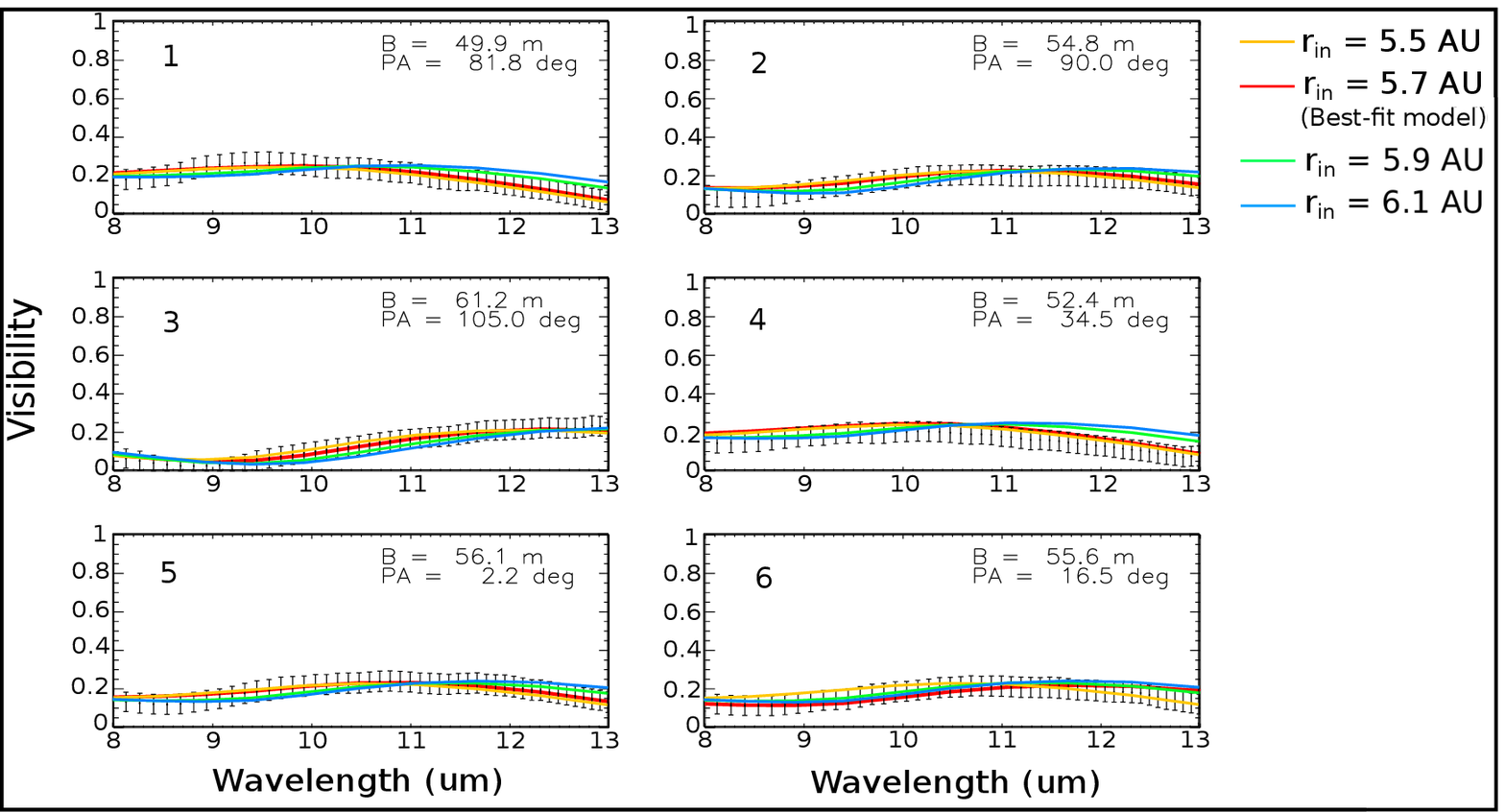}
 \includegraphics[width=90mm,height=60mm]{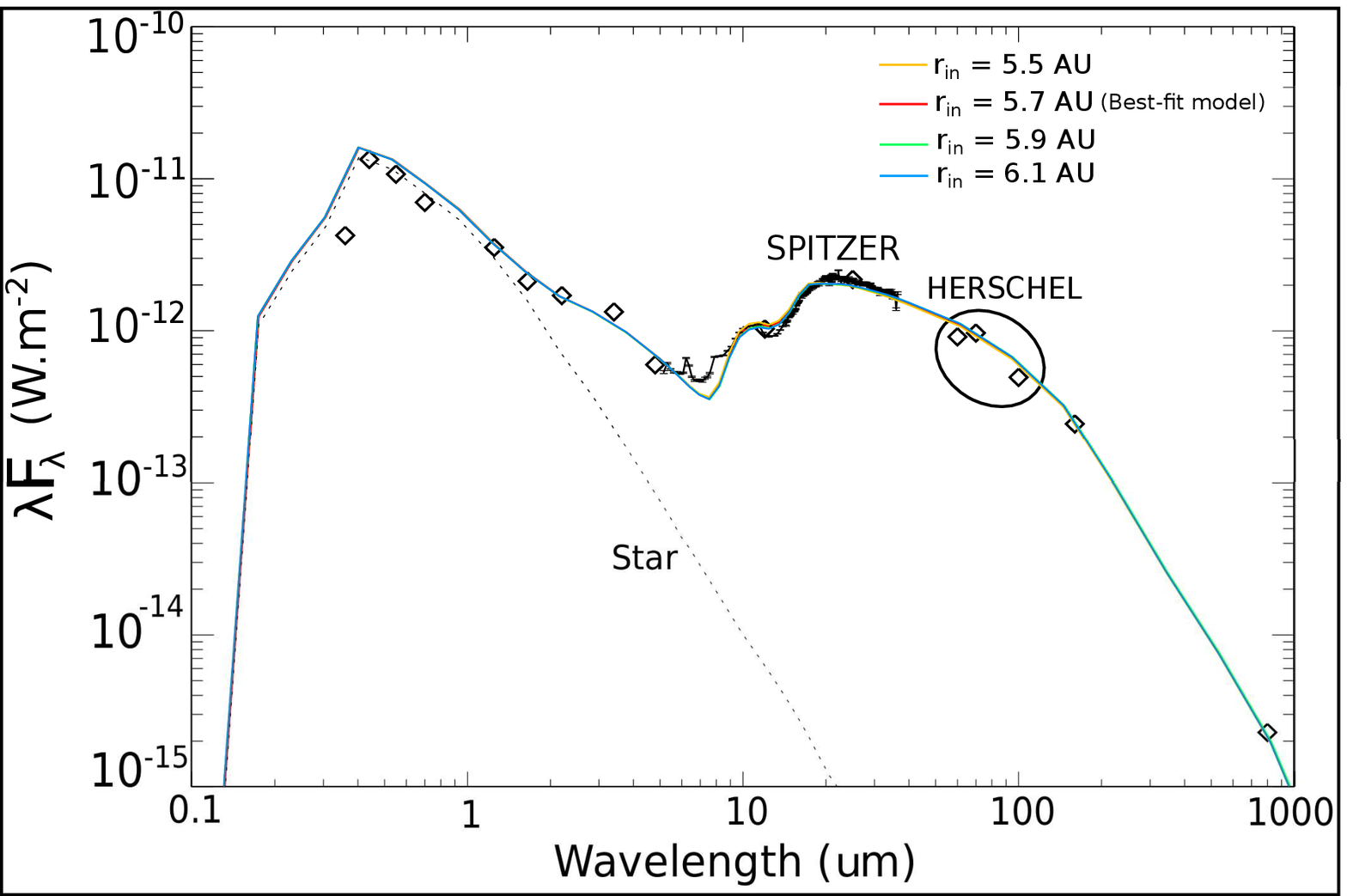}}
  \caption{{\footnotesize Same as in Fig.~\ref{fig:rin} but focusing on the effect on a change of the inner radius of the outer disk $r_{\rm in}$.}}
 \label{fig:p}
\end{figure*}
Finally, we show in Fig.~\ref{fig:h} the effect of varying the inner radius of the outer disk $r_{\rm in}$ on the MIR visibilities, which are directly probing this region of the HD~139614 disk. Indeed, it appears that a deviation by more than $\sim 0.3$~au, which is the 1-$\sigma$ uncertainty on $r_{\rm in}$ (see Table~\ref{tab:global}), induces a noticeable shift  in the sine-like modulation in the modeled MIR visibilities. The latter are no longer consistent with the measured MIDI visibilities, although the visibility level remains similar. No significant effect can be seen in the MIR SED.

\end{appendix}

\end{document}